# A novel clock and timing approach for achieving 200+ km ALMA baselines

*Brent Carlson, Herzberg Astronomy and Astrophysics, National Research Council Canada*
*(Brent.Carlson@nrc-cnrc.gc.ca)*

*Rev. 1, 2021-04-09*

## Abstract

*"A topic of particular interest to the NA ALMA Development Program" in the ALMA Cycle 9 Call for Project Proposals is longer baselines "to enable qualitatively new science." Research into the novel approach of "incoherent clocking" (IC) has been underway by the author at NRC-Penticton for the last ~2.5 years and this approach is poised to deliver this capability. In this approach, each antenna is outfitted with its own free-running, independent Local Oscillator (LO)—and thus is incoherent with all other antennas—and it is used as the antenna's local frequency reference to develop all down-conversion and digitization frequencies. Subsequently, using all-digital methods and all-digital fiber optics, both implemented with Commercial Off-The-Shelf (COTS) devices, the antenna LO's (aLO) varying frequency vs time is measured, relative to a central/common reference frequency, and the digitized science data is subsequently digitally corrected to the common reference before correlation and beamforming. The key advantage of this method for ALMA is a substantial increase in baseline length to 200 km or more (meaning fiber reaches of 100 km or more), at the lowest cost possible in terms of equipment and fiber routing for these longer baselines. A second advantage of IC is that the number of sub-arrays is no longer restricted by the number of photonic references—each antenna's observing frequency can be set independently of others, limited only by the agility of frequency synthesizers available at the antenna. Thus, the number of sub-arrays is limited only by the number of antennas available and, of course, by correlator agility in this regard. Finally, a subtle but important advantage of IC is that spurs and low-level spectral copies, due to interleaving in digitizers, do not correlate since they are digitizer clock frequency-dependent and therefore are at different frequencies for each antenna. This leads to better fidelity in visibility data products. What follows in this memo is a deep dive into IC research thus far, to hopefully convince the reader that the IC approach is indeed feasible for ALMA, even for Band 10.*

National Research Council Canada　　Conseil national de recherches Canada

## Document History

| Date | Revision | Person | Change Description |
|---|---|---|---|
| 2021-02-14 | A | B. Carlson | Initial release for internal review; Abbr & acronyms TBD. References still need to be fleshed out and cleaned up. |
| 2021-02-15 | B | B. Carlson | First completed version ready for internal review. |
| 2021-02-22 | C | B. Carlson | Fix some errors: PLL phase noise spurs; notion of ReSampler integrated with the Talon VCC OSPPFB; resampling error sensitivity loss. Still needs grammar fixes. |
| 2021-03-02 | D | B. Carlson | Simplify and clean up the parts of section 3 and section 4.1. Establish an aLO and Frequency Tracker phase noise budget of 20 fsec RMS based on current ALMA antenna electronics budget of 65 fsec. Still need to fill in some simulation results in section 4.1 |
| 2021-03-03 | E | B. Carlson | Fill in numbers for simulation results in section 4.1. |
| 2021-03-05 | F | B. Carlson | Change the ADC jitter equation on p. 14 to $2^{-(ENOBs+2.5)}$ to be more accurate. This puts the ADC ADEV at ~1.3e-11. |
| 2021-04-09 | 1 | B. Carlson | Change the ReSampler diagram Phase DDS and description to not require multiple DDDs, just one 'Phase_o' DDS, whose output is scaled to the required sky frequency before application to the I/Q mixer. Also, don't need multiple antenna synthesizers, just agile ones for desired tuning flexibility. Internally reviewed by SH and MP, first full release. Also, add Appendex with derivation of equation (5) on p. 29. |

# Table of Contents

National Research    Conseil national
Council Canada       de recherches Canada

Canada

# Table of Figures

**National Research Council Canada**  **Conseil national de recherches Canada**

Canada

## List of Abbreviations and Acronyms

1PPS – 1 Pulse Per Second.

3R – Re-timing, Re-shaping, Re-amplifying.

ADEV – Allan Deviation.

ADC – Analog-to-Digital Converter.

aLO – Antenna LO, free-running, independent of others.

AOS – Array Operations Site.

ASIC – Application Specific Integrated Circuit.

BiDi – Bi-Directional data transport on a single fiber, using two different wavelengths.

CDR – Clock Data Recovery.

cLO – Central LO, the reference clock source for the array.

COTS – Commercial Off-The-Shelf.

CRC – Cyclic Redundancy Check.

CTE – Coefficient of Thermal Expansion.

CTR – Central Timing Reference.

DAC – Digital-to-Analog Converter.

DCTD – Digital Central Timing Distributor.

DDS – Direct Digital Synthesizer. A "point-slope" linear synthesizer, where the "point" is phase and "slope" is "pinc" (phase increment.)

DTS – Data Transmission System.

DWDM – Dense Wavelength Division Multiplexing.

ENOB – Effective Number of Bits.

FIFO – First-In First-Out.

FIR – Finite Impulse Response (filter.)

FPGA – Field Programmable Gate Array.

HDL – Hardware Description Language.

IC – Incoherent Clocking.

LO – Local Oscillator.

LOS – Loss Of Signal.

**National Research Council Canada**    **Conseil national de recherches Canada**

Canada

LPFF – Low-Pass FIR Filter.  Refers to the filter after the Frequency Tracker.

LSB – Least Significant Bit.

MT – Message Type (of the IC_telem protocol.)

OCXO – Oven-Controlled Crystal Oscillator.

OEO – Optical-Electrical-Optical.

OSF – Observation Support Facility.

PID – Proportional, Integral, Differential.

PLL – Phase Locked Loop.

ReS – Short form for "ReSampler."

ReSampler – Refers to the entire ReS FIFO, interpolator, phase rotator processing block, including Re-sampling and Phase DDSs.

RFI – Radio Frequency Interference.

RFSoC – RF System on a Chip.

RT – Round Trip.

RTL – Register Transfer Level.  A subset of a HDL, which can be compiled ("synthesized") into FPGA hardware.

rtm – Round-trip multiplier.

SB – Sync Bit (of the IC_telem protocol.)

SERDES – Serializer/De-serializer.  Include a CDR PLL.

SEU – Single Event Upset.

SFP – Small Form factor Pluggable.

SOF – Start of Frame (of the IC_telem protocol.)

SSR – Super Sample Rate.

TE – Timing Epoch; for ALMA this marks every 48 msec epoch, coincident with a 1PPS every 6.0 seconds.

TIC – Time Interval Counter.

TL – Timing Link.

Tracer (tracer) – complex tone derived from DDS driven by an appropriate clock.  Typically this is ~10 MHz and 16-24-bits in precision.

VLBI – Very Long Baseline Interferometry.

National Research   Conseil national
Council Canada     de recherches Canada

Canada

# 1   Introduction

Extending ALMA baselines is a key topic of interest to the NA ALMA Development Program, Cycle 9 Call for Proposals.  This interest has prompted the author to write this memo to explore and explain, in considerable detail, the research into "incoherent clocking" (IC) that has been underway at NRC-Penticton over the last nearly 2.5 years.  This memo is therefore a key applicable document for a Proposal submission,  in response to the Cycle 9 call, to perform an experiment to empirically verify that this method can be used for ALMA, even for Band 10 over 100+ km of aerial fiber.

In this memo, in addition to explaining the research, a number of architectures that may be used for ALMA are developed and described.  These are developed within the author's understanding of ALMA infrastructure and informed by descriptions of such contained in [1], including maximum baselines possible due to geographical and geopolitical considerations.  The "200+ km baselines", requiring fiber lengths of 100+ km ignores these issues, instead indicating what the technology is capable of doing.

The first notion of IC is described in [2].  Since then, much progress has been made in refining the approach, leading to an all-digital implementation.  Of course, "all-digital" does not eliminate the need for an appropriately low phase noise LO reference at the antenna, multiplied up by, and used in, the RF continuous-time domain until the signal is finally digitized in the ADC.  "All-digital" therefore refers to all processing performed after this point, including sufficiently accurately measuring the antenna LO ("aLO") frequency vs time relative to a common reference LO ("cLO"), e.g. the H-maser, over digital fiber optics connections and, using these measurements, digitally correcting the digitized science data into the common reference domain before correlation and beamforming.

The ability to do these measurements and apply corrections using all-digital methods and processing seems like a fantastic claim and at this point the reader might be tempted to stop reading.  However, in this memo the approach, research results, applicability to ALMA, and risk analysis is done to hopefully convince the reader that the probability that the IC method will *not* work for ALMA is very low and may even be zero.

# 2   Overview

The IC approach is a digital method for providing a clock and timing solution for any scale and frequency of radio telescope.  It uses a "measure and correct" approach wherein each antenna operates using its own independent LO (aLO), its frequency vs time is measured at a central site relative to a common reference cLO, and then the digitized data is digitally "corrected" (interpolated and phase rotated) before correlation and beamforming.

Since the aLO is measured in a common reference clock domain at a central site, naturally aLO's frequency—or more likely a frequency derived from, and locked to, it—must be transmitted there via a media such as fiber, whose delay is changing with time.  Thus, a round-trip measurement is performed to remove this temporal delay uncertainty from the frequency measurement and, additionally, digital filtering is used is remove temporal round-trip measurements on timescales less than some factor times the round-trip delay.  Since clock steering is not occurring, what this means is that a longer fiber requires a more stable aLO, however as will be shown, it is always the case that *any aLO frequency measurement variations on timescales that are a priori known to not be intrinsically in the aLO can be digitally filtered out*.  This affords considerable performance improvements over clock steering approaches, allowing virtually any baseline extent to be accommodated, with the limitation only being the stability of the aLO for the required (highest) observing frequency and fiber reach.

Additionally, digital methods are used for all measurements and to convey a frequency derived from the aLO to the central site. Thus, all measurements and corrections occur in digital devices such as FPGAs, and all signals flowing between the antenna and the central site use COTS digital fiber optic transceiver modules and lowest-cost aerial fiber routing methods. There is, of course, a continuous-time aspect to this approach since it would be impossible to measure aLO's varying frequency using strictly discrete-time processing. Thus, there is something of a "cross-over" between continuous-time and discrete-time in the IC approach, but all-digital methods are nevertheless used.

A greatly simplified block diagram of the IC method is shown in Figure 2-1 for a Direct Digital Conversion (DDC) receiver; for a heterodyne receiver such as ALMA one or more down-conversion stages in front of the ADC/digitizer is of course needed.

In this diagram, timing information is conveyed over a full-duplex digital fiber optic "Timing Link" (TL) here containing the digitized data as well, by clocking the serial transmitter with a derivative of the aLO, recovering the clock at the central end with a Clock-Data Recovery (CDR) PLL[1], and then measuring that recovered clock's frequency in the central common (reference) clock domain cLO. This is a typical case, but a link separate from the digitized science data could be used for the TL and indeed, the TL could be wireless (as of course could be the link for the digitized data.)

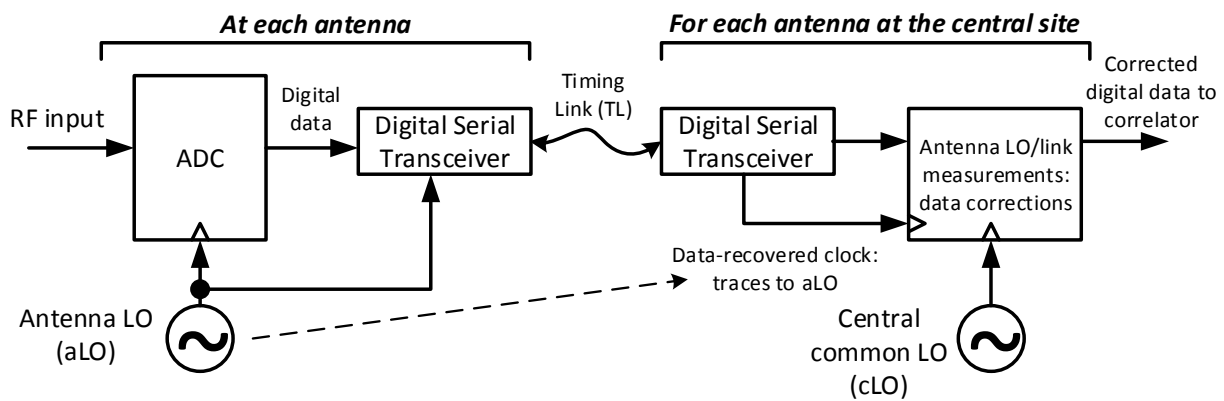

*Figure 2-1  Simplified block diagram of the IC approach.*

A more detailed functional block diagram of a typical IC implementation is shown in Figure 2-2 (included here; a more detailed signal flow description can be found in [3].)

In this diagram:

- All down-conversion and digitization clocks are derived from the[2] aLO. Not shown are all of the required derived frequencies for these purposes, just that ultimately all are derived from the

---

same aLO, and that any differential phase noise present in these derivations are inconsequential[3].

- The digitized science data as well as "Round-trip measured phase & delay + tracer signaling" are integrated onto the same fiber and transmitted to the central end using SERDES and (COTS) digital fiber optic transceivers. Any signal repeating between the antenna and the central end must be "transparent", meaning that timing information is conveyed through them. These repeaters may be "3R" (Re-timing, Re-shaping, Re-amplifying) OEO type or OO type; jitter cleaning may occur, but it must not clean jitter (phase/frequency wander) that is intrinsic in the aLO on timescales that must be measured.

- At the central end, the received signal is looped back and transmitted back to the antenna end. The return fiber/repeater dynamic behaviour, on time scales ≥ the phase/frequency wander of the aLO that must be measured and corrected, must be sufficiently well matched to the incoming direction such that the single-direction phase is simply the round-trip phase measurement divided by 2[4].

- The "tracer" is a high dynamic range complex digital tone, derived from the receive SERDES CDR PLL clock at the central end and antenna (loopback) end as required. Its phase and frequency therefore changes in precise proportion to CDR PLL clock phase and frequency changes. For example, for a 10G (Gbps) SERDES & digital fiber optic link, the CDR PLL clock present in the FPGA fabric is typically 250 MHz; this 250 MHz clock drives a 16-bit free-running Direct Digital Synthesizer (DDS), with a "pinc" value, typically a prime (relative, to the number of phase steps) number, chosen so that subsequent sampling of the DDS's digital phase across digital clock domains[5] doesn't beat with any clock frequency and that the tracer frequency is sufficient for phase and frequency measurement performance, typically ~10 MHz. Also, since tracer phase is converted to a complex tone after phase sampling into the required clock domain, it has perfect quadraturity[6], essential for accurate and precise phase and frequency measurements. Finally, the tracer tone is not in itself transmitted across the TL fiber, rather signaling to maintain synchronization of tracer DDSs at various places, is transmitted instead. This greatly decreases the digital bandwidth needed for it.

- At the antenna end, the phase of the returned tracer is continuously measured against the phase of the outgoing tracer and this information is transmitted to the central site on the same fiber optic link as the science data and tracer signaling. Typically, a new 32-bit phase measurement is made and transmitted every ~10 μsec and no attempt is made to digitally filter these measurements since such filtering is provided by the post-Frequency Tracker Low-Pass FIR Filter (LPFF.) Additionally, the round-trip delay is unambiguously measured using a time-interval counter (TIC, clocked by an aLO-derived frequency) between a pulse at the transmitter, and its return, with the measured value also transmitted to the central site on the same link. This round-trip delay provides a sufficiently accurate measurement of the round-trip (and therefore single-direction) delay of the fiber, needed for time-epoch marking of antenna digitized science data (1PPS and/or 48 msec TE) and to facilitate accurate Walsh phase change setting at the antenna (see subsequent discussions of these and associated issues for further details.)

---

[3] Meaning that it doesn't need to be measured and corrected. If it does, then the derived frequency that drives the digital SERDES transmitter must be derived from an appropriate point to capture that information.

[4] There is some possibility, TBC, that some mismatch may be calibrated-out by analysing the central-end Frequency Tracker PID servo Phase_err variable (Figure 4-2) and adjusting the division factor of 2 appropriately to minimize its level.

[5] Using gray-code and memory methods to ensure no code errors. Discrete-time digital phase sampling also generates phase noise, most of which is filtered out with the post-Frequency Tracker Low Pass FIR Filter (LPFF.)

[6] Meaning that the in-phase (cos) and quadrature (sin) components are 90 deg out of phase, albeit with noise proportional to the numerical precision used, but with no net bias.

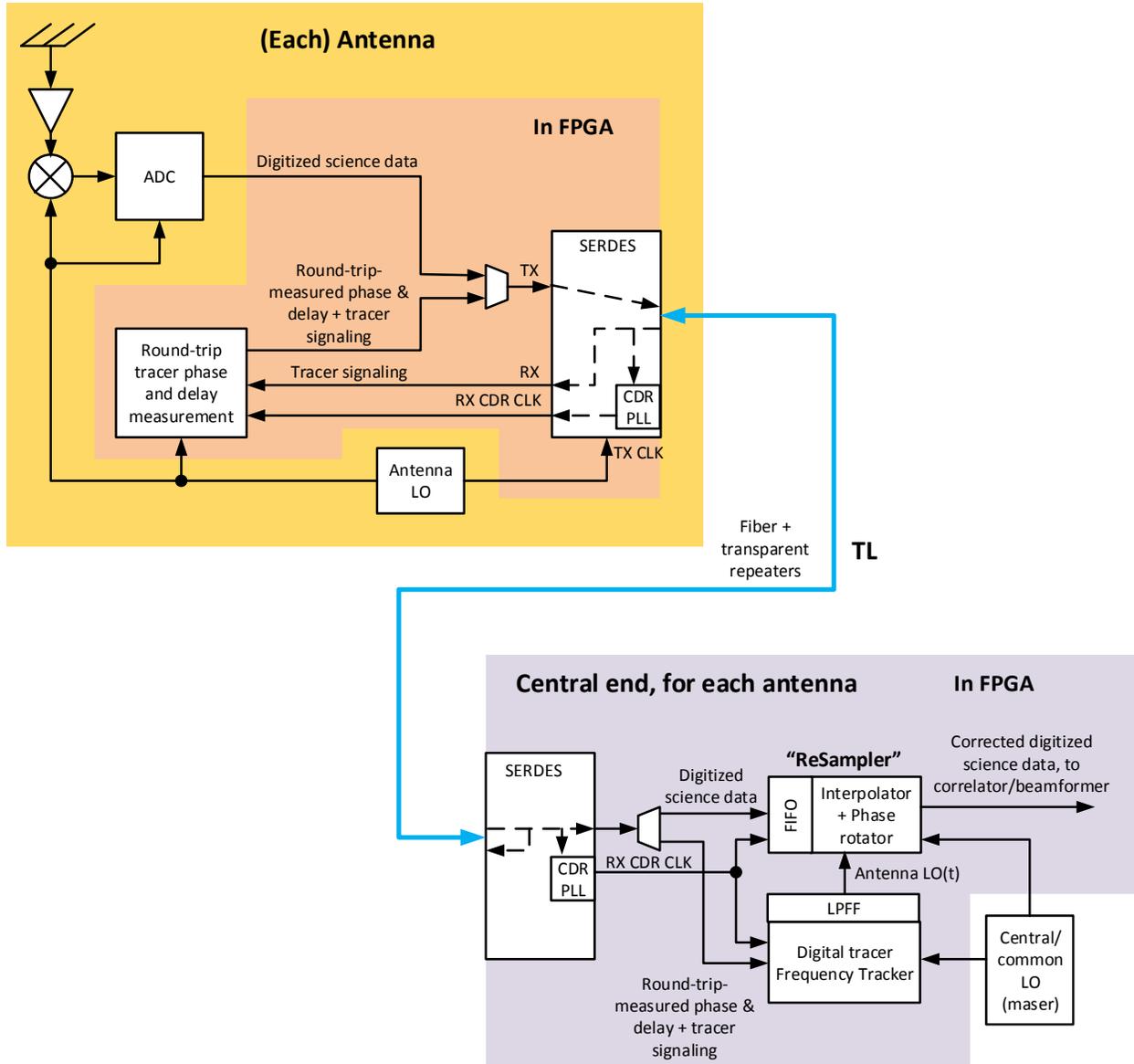

*Figure 2-2  Typical IC implementation functional block diagram.*

- At the central site, the round-trip-measured phase is applied as a compensating factor in the (digital) Frequency Tracker, which is continuously measuring the tracer frequency at that point, largely removing fiber delay variations from these measurements. As will be shown, only the round-trip measured phase on time scales ≥ aLO phase/frequency changes that must be measured, actually have an effect on the net after-LPFF result.

- Raw Frequency Tracker measurements, typically one every ~10 μsec, are filtered using the LPFF. This filter, whose DC gain[7] must be precisely known, removes any measured frequency measurement variations on timescales ≥ aLO phase/frequency variations that must be corrected. Also, as mentioned, the filter also removes the vast majority of phase noise due to digital phase sampling, fiber delay perturbation effects known not to be in the aLO, and FPGA clock jitter, which is typically 100 to 350 ps pk-pk.
- The LPFF output frequency measurements drive DDSs, which drive the ReSampler (see Figure 4-17.) All digital mathematical operations are sufficiently accurate and unbiased since there is no feedback to determine if the applied corrections are "correct." Such accuracy and unbiased operation can be obtained with, for example, the use of well-known unbiased digital rounding operations.
- The digitized science data output from the ReSampler is now fully-corrected to the reference clock domain, as are similar outputs from other antennas, ready for correlation and beamforming.

# 3   Antenna clock stability and LPFF requirements

As mentioned, in the IC approach, a longer fiber requires a more stable aLO. This is specified in terms of its Allan Deviation (ADEV) using the well-known ADEV equation relating the frequency under consideration, its frequency change, and phase change over a time duration Tau. This is discussed in [3] and presented in the following equations:

$$AD\_c \leq \frac{\varphi_{RMS} \cdot v_{link}}{2L \cdot rtm \cdot fm \cdot v} \qquad (1)$$

and:

$$Tau\_c \cong \frac{2L}{v_{link}} \cdot rtm \cdot fm \qquad (2)$$

Where:

AD_c is the aLO Allan Deviation at Tau critical, "Tau_c", according to Figure 3-1.

$\varphi_{RMS}$ is the allowed RMS phase noise.

$v_{link}$ is the velocity of signal propagation through the fiber; for single mode it is ~2/3 c.

L is the fiber length.

$v$ is the frequency being considered.

---

[7] At frequencies other than DC there will, of course, be some gain variation in any LPFF filter. However these gain variations affect only tracer frequency variations, which are very small. A full analysis of required LPFF bandpass ripple requirements for a given aLO ADEV, digitizer frequency, and sky frequency is TBD, noting that the number of LPFF taps can be increased to achieve very low bandpass ripple, typically <0.001 dB, and that no noticeable effects of this kind of bandpass ripple have been found in any testing to date. For example, 0.001 dB gain error at f_tracer_er of 1e-4 Hz, results in an error of 2e-8 Hz. Even for Tau=10 sec, this is phase error of 1e-7 cycles.

rtm is the round-trip multiplier—fiber delay changes on timescales short than this factor times the round-trip delay are not reliable and need to be filtered out.

fm is the "filter multiplier" and accounts for the delay of the LPFF; for the case where the FIFO in front of the ReSampler ~matches the delay of the LPFF, fm=1; when not, it is typically ~15.

Total antenna phase noise including the aLO, its frequency synthesizers, mixers & digitizers, the analogue signal, as well as measurement of the aLO by the central-site Frequency Tracker[8] must be taken into account. From Table 1 [4], the total phase noise for the current electronics is 65 fsec ~= 23 deg at 1 THz. In that document, however, an allocation for the antenna LO, as delivered by the round-trip phase-corrected method employed at the Warm Cartridge Assembly (WCA) ingress point [21], could not be found. Therefore a reasonable allocation of ~20 fsec RMS (~8 deg RMS at 1 THz) is chosen here as a minimum phase noise performance requirement for the aLO (Tau < Tau_c) and the Frequency Tracker (Tau > Tau_c), since 20 fsec increases the total electronics phase noise to ~68 fsec, a negligible increase.

***Example:***

*For ALMA mixers/downconverters at v=1 THz, L=100 km, $\varphi_{RMS}$=8 deg, and for rtm=10 [5] leads to an aLO AD_c~=2e-12 @ Tau_c~=10 msec.*

*For ALMA 2030 digitizers @40 Gs/s, 5 ENOBs, the allowable phase noise of the aLO (consistent with, for example, AD9213 data sheet, Figure 90) is $\sim 2^{-(ENOBs+2.5)}\sim$= 5e-3 cycles RMS at 40 GHz and therefore AD_c~=1.4e-11, also at Tau_c~=10 msec.*

Thus, the aLO ADEV requirement for the 1 THz downconverter is the governing value. Of course, the antennas that are much closer to the central site, i.e. L<<100 km, would have a less-stringent requirement, many by at least an order of magnitude, but such depends on digital Frequency Tracker performance as we shall see in section 4.1.

If the ReSampler FIFO depth does not match the LPFF delay, for the ALMA example above at 1 THz with fm=15, the aLO AD_c~=1e-13 @Tau_c~=0.15 sec, requiring a device that may be difficult and costly to obtain.

---

[8] It is important to point out that aLO and Frequency Tracker phase noise contributions are mutually-exclusive—for timescales Tau < Tau_c the Frequency Tracker is essentially turned off if fm=1 (with $f_{c\_LPFF}$ according to equation (3)); for timescales Tau > Tau_c, aLO phase noise is completely compensated to the precision of the Frequency Tracker.

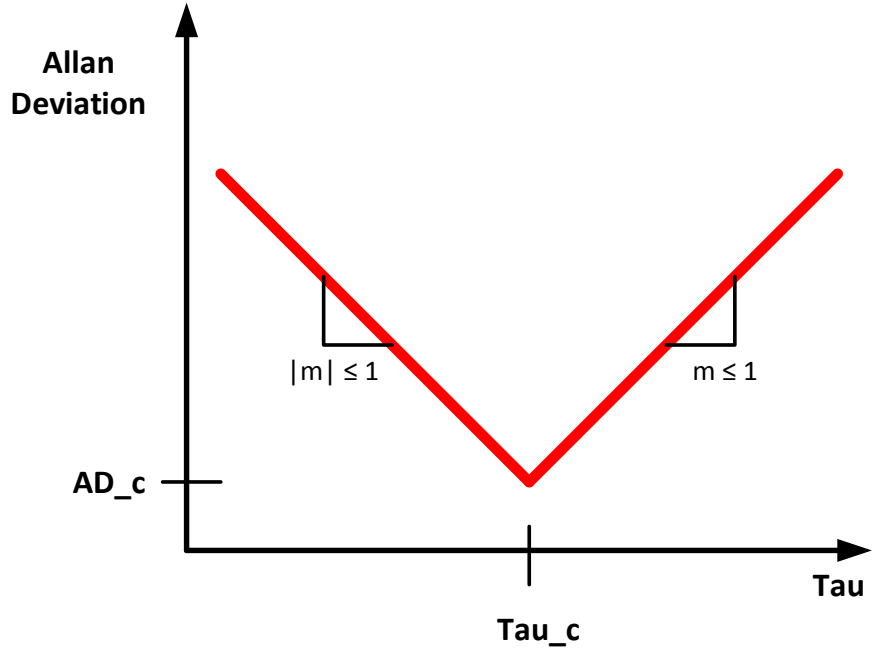

*Figure 3-1 Log/log IC Allan Deviation curve. The ADEV of aLO must be below this curve. On timescales < Tau_c, aLO phase/frequency noise/wander is not measured or corrected; on timescales > Tau_c, aLO phase/frequency wander is measured and corrected (re-sampled, phase rotated.) Not shown is that for Tau >> Tau_c, of course, the curve bends over to the limit of aLO absolute frequency tolerance.*

To minimize aLO stability requirements, fm=1 and therefore the ReSampler FIFO requires memory to buffer the digitized science data to match the LPFF delay, $\tau_{LPFF}$, which can be significant. In order to remove all aLO frequency variations that are not intrinsically in the aLO, we take the cut-off frequency of the LPFF, $f_{c\_LPFF}$, as:

$$f_{c\_LPFF} \cong \frac{1}{4Tau\_c} \qquad (3)$$

The delay through the LPFF, for ~80 dB reject band attenuation and very low passband ripple, has been found to be:

$$\tau_{LPFF} \cong \frac{3}{f_{c\_LPFF}} \qquad (4),$$

with a ~100 dB reject band filter, ~15% longer.

For the example above, at Tau_c~=10 msec, $f_{c\_LPFF}$~=1/0.04sec=25 Hz and $\tau_{LPFF}$~=3/$f_{c\_LPFF}$~=0.12 sec. For a 40 Gs/s stream at 5b/sample, a total of 3 GBytes *per stream* is therefore required for the ReSampler FIFO.

# 4   Technical descriptions

## 4.1   Central-site Frequency Tracker design and performance

The previous section provides the theory to establish aLO stability, $f_{c\_LPFF}$, and ReSampler FIFO memory requirements. However, a key part of feasibility analysis for any IC application is the design and performance of the central-site Frequency Tracker since its performance is a limiting factor. Thus, this topic is addressed in considerable detail in this section.

Although it is sufficient for the aLO's ADEV to be below the curve of Figure 3-1, doing so, for the extreme ALMA application of $v=1$ THz, may require digital logic resources and performance (i.e. tracer DDS clock rate, phase sampling rate, and Frequency Tracker clock rate) to meet the demand of tracking the highest df/dt of the tracer present. If the aLO's ADEV follows the curve of Figure 3-1 for Tau > Tau_c, df/dt is the same for all Tau > Tau_c. However, if the aLO curve is more along the lines of Figure 4-1, then df/dt is relaxed, requiring fewer digital logic resources and less digital logic performance, since it is at Tau_c where tracking frequency variations needs to start, not where it must be fully doing so. This kind of curve is normally the Allan Deviation of a crystal oscillator anyway, rather than the curve of Figure 3-1.

Actual logic resources and performance may be determined empirically, and the cost (and power) thereof balanced with the cost of the aLO. As will be seen, there is some room for Frequency Tracker performance improvement resulting in lower phase noise and higher df/dt tracking, although pushing the limit does come with increased cost and power, something one should always strive to minimize.

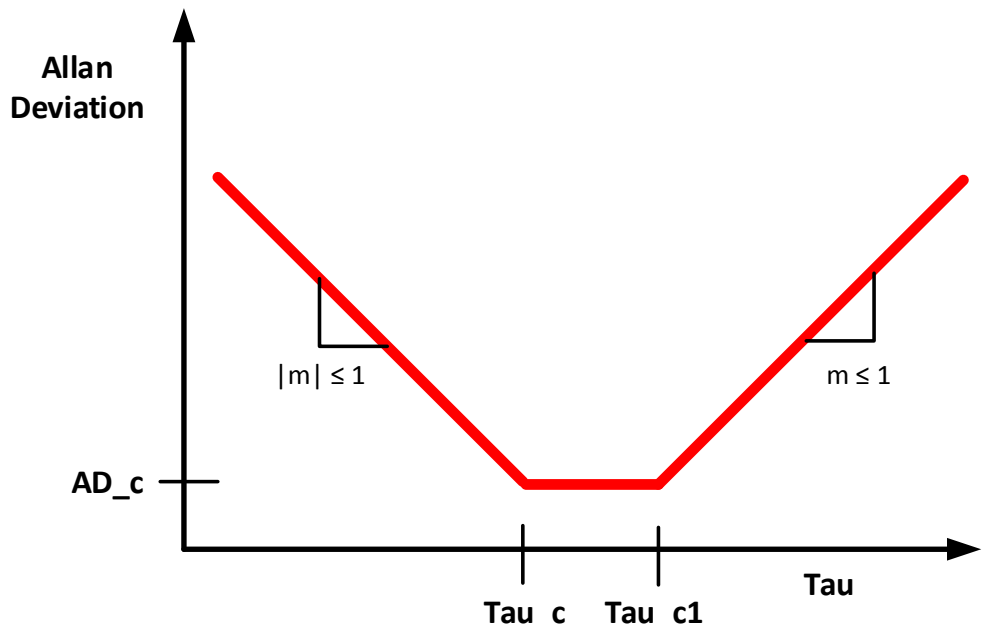

*Figure 4-1 Antenna LO log/log ADEV curve requiring less logic performance in the Frequency Tracker. Frequency Tracking starts to occur at Tau_c, but isn't at full performance until Tau_c1, which is typically 5x to 10x Tau_c.*

A block diagram of the Frequency Tracker (i.e. one possible design of the Frequency Tracker) is shown in Figure 4-2. Its operation is fundamentally very simple in concept, although not shown in the diagram is the number of bits carried and used for each operation, which vary from 16b to 64b.

The phase detector is the short-term-accumulated complex-conjugate cross-correlation of a high dynamic range tone from the tracer DDS, driven by the received data Rx CDR PLL-recovered clock, with the feedback tracer DDS' high dynamic range tone, driven in the reference clock domain.

The output of the phase detector drives a phase PID calculator, whose output is an accumulated phase correction to the feedback DDS phase input on each iteration of the loop.

The phase correction above drives a frequency PID calculator, whose output f_tracer_loop is an accumulated frequency correction to the feedback DDS pinc (phase increment) register. Nf f_tracer_loop measurements are averaged to produce f_tracer_raw, which goes to the LPFF for filtering.

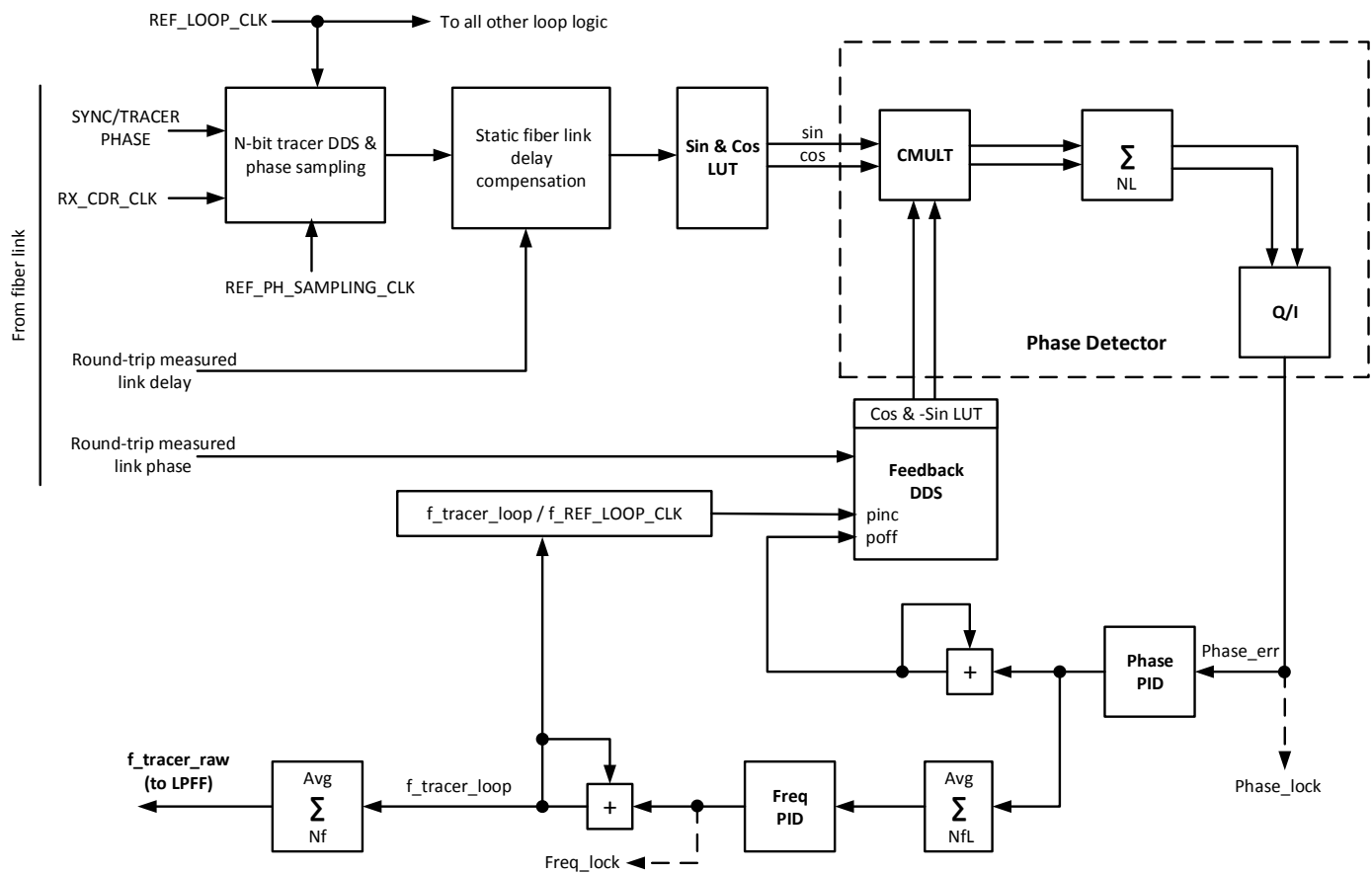

*Figure 4-2 Frequency Tracker block diagram. Phase_lock and Freq_lock are asserted when the variation in loop signals at the indicated points are sufficiently low.*

The phase PID and frequency PIDs are assigned their own P, I, and D gains and (normally) the former's gains are much higher than the latter's. This means that the phase PID (inner loop) is always phase-locked and tracking the Rx CDR PLL clock-driven tracer DDS, and the frequency PID (outer loop) tracks its frequency changes[9]. For the kind of phase jitter/wander that the tracer Rx CDR PLL-driven DDS sees, for reasonable phase PID gains, phase-lock is always achieved and assured. It is the frequency PID gain coefficients that determine df/dt tracking and resulting tracer frequency output vs time that is required to meet a given aLO Allan Deviation at the required phase noise for the required maximum frequency.

---

[9] It is possible to only have a phase PID loop, with the phase change on each iteration of the loop driving the ReSampler DDS. This option has not been explored in detail and may work just fine if the antenna LO's absolute frequency error is well-constrained.

For Frequency Tracker logic and performance presented in [3] (referring to Figure 4-2: N=16-bit tracer DDS operating at RX_CDR_CLK=250 MHz; phase sampling at REF_PH_SAMPLING_CLK=303.75 MHz; PID loop logic operating at REF_LOOP_CLK=101.25 MHz—phase-locked to 303.75 MHz; NL=51, NfL=4, Nf=4, phase PID gains[10]=$2^{-11}$, $2^{-23}$, $2^{-17}$; freq PID gains=$2^8$, $2^{-7}$, $2^7$; $f_{C\_LPFF}$~=25 Hz (-15 dB))[11] the best phase noise performance found for a ~3e-5 Hz pk sinusoidal tracer frequency variation (ADEV$_{aLO}$ of 2e-12 x 10 MHz x √2) at 2.5 Hz (i.e. Tau_c1=0.1 sec which is 10*Tau_c) is ~4.8e-7 cycles RMS at a tracer frequency of ~10 MHz, which translates to ~17 deg RMS at 1 THz, higher than the allocated limit of ~8 deg RMS previously established.

This result was found with a discrete-time MathCad model of the Frequency Tracker that has been found to be consistent with FPGA RTL[12] simulation results ([6], with MathCad-RTL comparison reported at [7].) The MathCad model is used for performance evaluation since it runs ~200-300X faster than an FPGA RTL simulation.

Plots from this simulation are shown in Figure 4-3 to Figure 4-7 below:

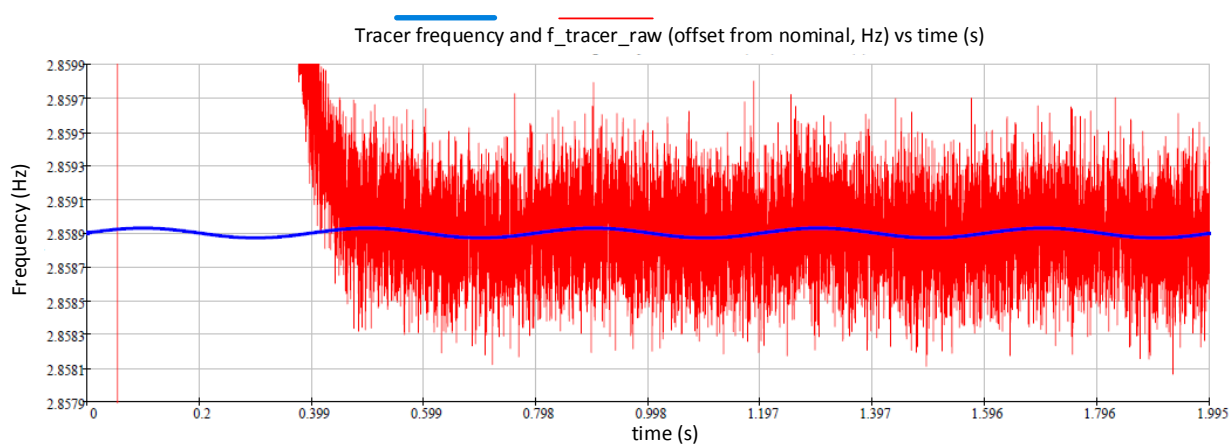

*Figure 4-3  Raw Frequency Tracker result before LPFF.  In this simulation no fiber link jitter is introduced since the effect of such is well understood [6][8][3].  Here, the tracer frequency is ~10 MHz; it has a constant frequency offset (from nominal) of ~2.86 Hz and a sinusoidal frequency variation of ~3e-5 Hz pk at 2.5 Hz.*

---

[10] Powers of 2 are used since application of gain entails simple bit shifts.  This simplification has limitations in gain selection, although such was not found to be a problem.

[11] As well, using the specified Intel Stratix-10 FPGA I/O PLL clock jitter of 175 psec pk-pk max, assuming a Gaussian distribution and 2.5σ (a reasonable assumption, consistent with the confidentially-obtained "Intel Stratix 10 I/O PLL Validation Report, May 2018"), noting that in the simulations FPGA clock jitter has a much smaller effect on phase noise than discrete-time phase sampling.

[12] RTL="Register Transfer Level."  This is FPGA HDL (Hardware Description Language) code that is "synthesizable" and maps to actual FPGA hardware.  A "simulation" of the RTL code precisely indicates how the FPGA will perform as long as timing closure is achieved (i.e. when achieved, it is guaranteed to run at the required speed(s).)

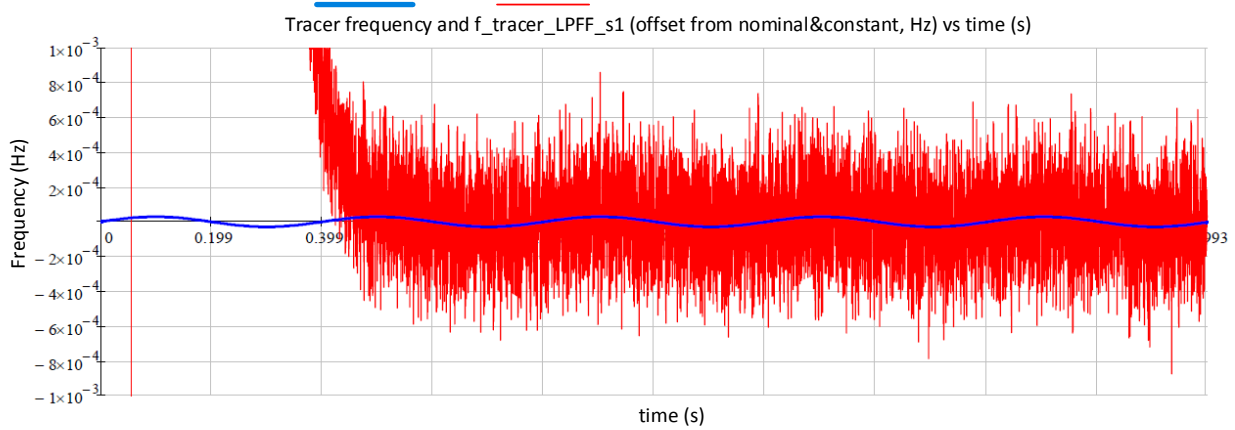

*Figure 4-4  Frequency Tracker result after stage 1 of the LPFF with $f_{c\_LPFF\_s1}$~=3 kHz.  Here, the constant offset of ~2.86 Hz in Figure 4-3 has been removed from the result to more clearly see the imposed sinusoidal frequency variation due to antenna clock instability.  In this case the RMS phase error is ~5e-7 cycles of phase at f_tracer~=10 MHz.*

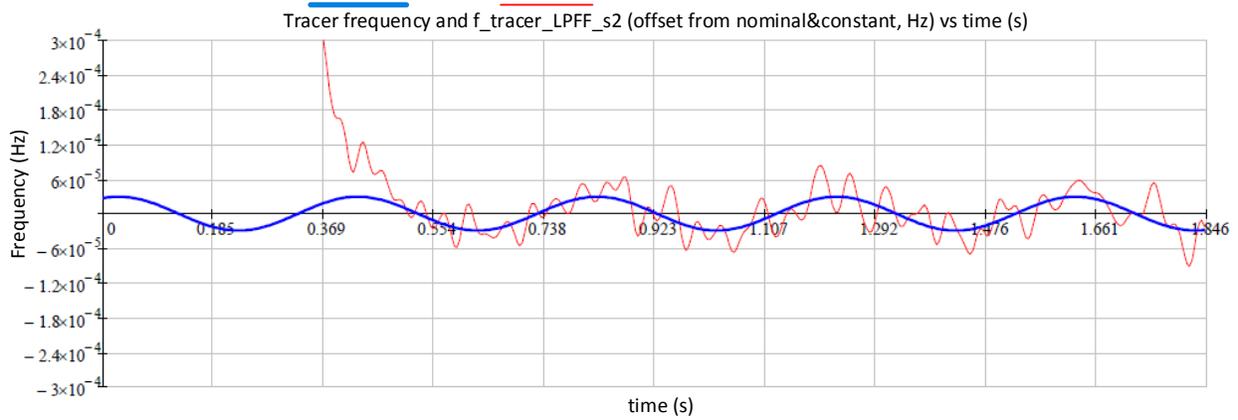

*Figure 4-5  Frequency Tracker result after stage 2 of the LPFF with $f_{c\_LPFF\_s2}$~=25 Hz.*

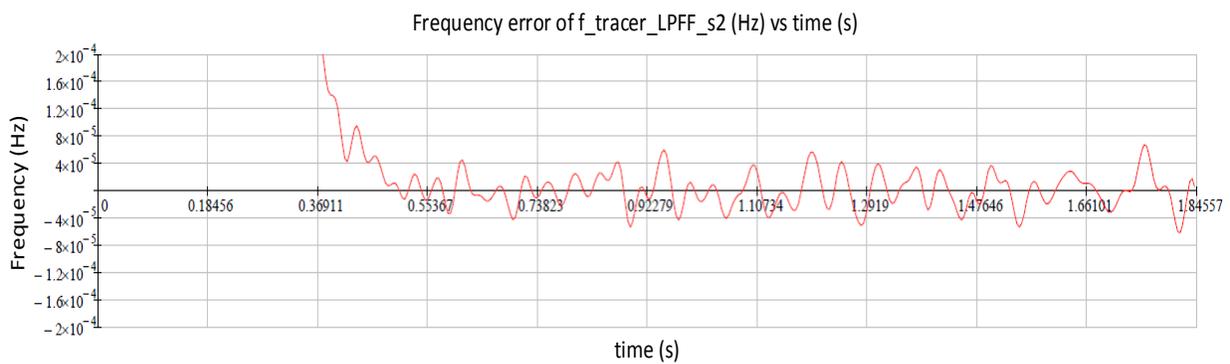

*Figure 4-6  Frequency Tracker frequency error after stage 2 of the LPFF with $f_{c\_LPFF\_s2}$~=25 Hz.*

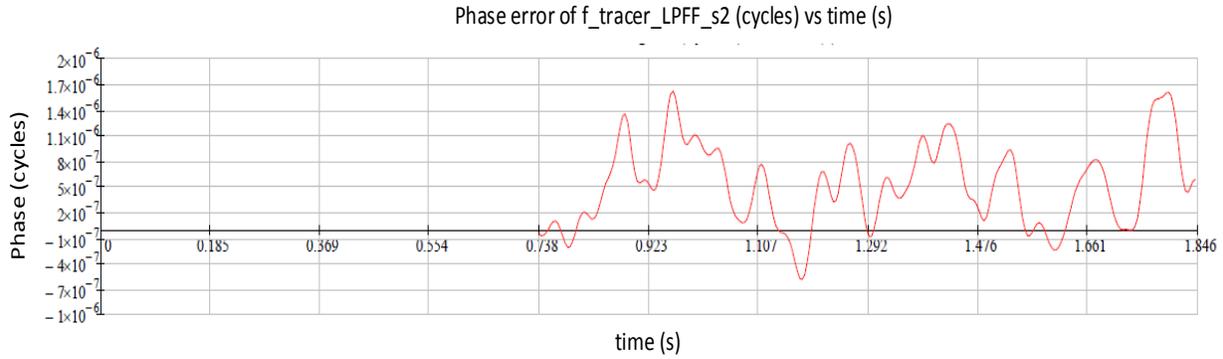

*Figure 4-7  Frequency Tracker phase error after stage 2 of the LPFF with $f_{c\_LPFF}$~=25 Hz.  The RMS phase error is ~4.8e-7 cycles at $f\_tracer$~=10 MHz, which translates to 17 deg RMS at 1 THz.*

From the above results it would seem that there is no utility for stage 2 of the LPFF since the RMS phase error out of each one is approximately the same.  However, stage 2 is absolutely essential to remove uncompensated fiber link jitter, not included in this simulation but reported in [6], [8], and [3].

Running the simulation with ADEV~2e-12, the tracer DDS operating at 500 MHz, sampled at 603.75 MHz, processed at 301.875 MHz, a tracer frequency of ~41 MHz , and a proportionally higher tracer frequency variation of ~1.2e-4 Hz pk at 2.5 Hz (with NL=51, NfL=16[13], Nf=4, phase PID gains=$2^{-11}$, $2^{-23}$, $2^{-17}$; freq PID gains=$2^{10}$, $2^{-5}$, $2^9$; $f_{c\_LPFF}$~=25 Hz) yields an RMS phase error (of the 41 MHz tracer) of ~1.6e-6 cycles, which translates to ~14 deg RMS at 1 THz.

This highlights the case (mentioned previously) where aLO stability requirements are driven by the performance of the Frequency Tracker (of course, coupled with maximum observing or digitization frequency), rather than fiber length L.  By reducing the aLO's ADEV, Frequency Tracker PID gains can be reduced, resulting in lower phase noise in the measured result.

For example, re-running the previous 10 MHz tracer simulation with an aLO ADEV~=5e-13, resulting in a 2.5 Hz sinusoidal frequency variation of ~7e-6 Hz pk, with phase PID gains=$2^{-11}$, $2^{-23}$, $2^{-17}$ and freq PID gains=$2^7$, $2^{-8}$, $2^6$ yields an RMS phase error of ~2.9e-7 cycles or ~10 deg RMS at 1 THz.  Re-running the above 41 MHz tracer simulation with a 2.5 Hz sinusoidal frequency variation of ~3e-5 Hz pk, with freq PID gains=$2^9$, $2^{-6}$, $2^8$ yields an RMS phase error of ~9.6e-7 cycles or ~8.4 deg RMS at 1 THz

If the aLO must be more stable for a longer Tau as above because of Frequency Tracker phase noise, it also means that the fiber length L can increase beyond the 100 km example used thus far, according to equation (1).  For the above curve of ADEV≤5e-13 @ Tau_c=10 msec and ADEV≤5e-13 @ Tau_c1=0.1 sec, full frequency tracking occurs at Tau~=0.1 sec; evaluation of equation (1) for an ADEV of 5e-13 and $\varphi_{RMS}$~=8 deg yields L~=440 km.

If extra fiber length L is not required, then the rtm factor of equations (1) and (2) can be increased by the same factor, in this case ~4x to rtm~=40, by decreasing $f_{c\_LPFF}$ (equation (3)) by the same factor, adding more assurance that fiber delay perturbations are completely removed via round-trip measured compensation and filtering with the LPFF.  Here it means that $f_{c\_LPFF}$~=6 Hz.  However, $\tau_{LPFF}$ then increases by that same factor to ~0.5 sec (equation (4)), which could introduce a latency problem for the ALMA

---

[13] Increased to 16 from the previous 4 to keep within MathCad's array size limitations.

VLBI tied-array feedback loop turn-around time of 1-2 seconds (6.5.10 of [1]), for ~200 km baselines. This leads to the notion that $f_{c\_LPFF}$ and ReSampler (Figure 4-17) FIFO delay can be dynamically configured for VLBI sub-arrays, which could be restricted to an aperture size such that $\tau_{LPFF}$ isn't an issue in this regard.

## 4.2   Round-trip phase tracker

A block diagram of the round-trip phase tracker circuit, present in the antenna FPGA, is shown in Figure 4-8.  Its operation is similar to the Frequency Tracker, except that here accumulated feedback phase is added to the round-trip ("RT") tracer DDS output to close the loop.  Then, accumulated feedback phase is averaged Np times, divided by 2, and sent to the central site on the same serial channel as tracer signaling to be used by the Frequency Tracker.  In the lab demo described in [3] NL=51, Np=4, and other clock frequencies are the same as the Frequency Tracker defined above.  This results in phase measurements being sent to the Frequency Tracker with a cadence of ~10 μsec.  Other than some averaging, phase measurements aren't filtered, instead they are effectively filtered by the LPFF after the Frequency Tracker.

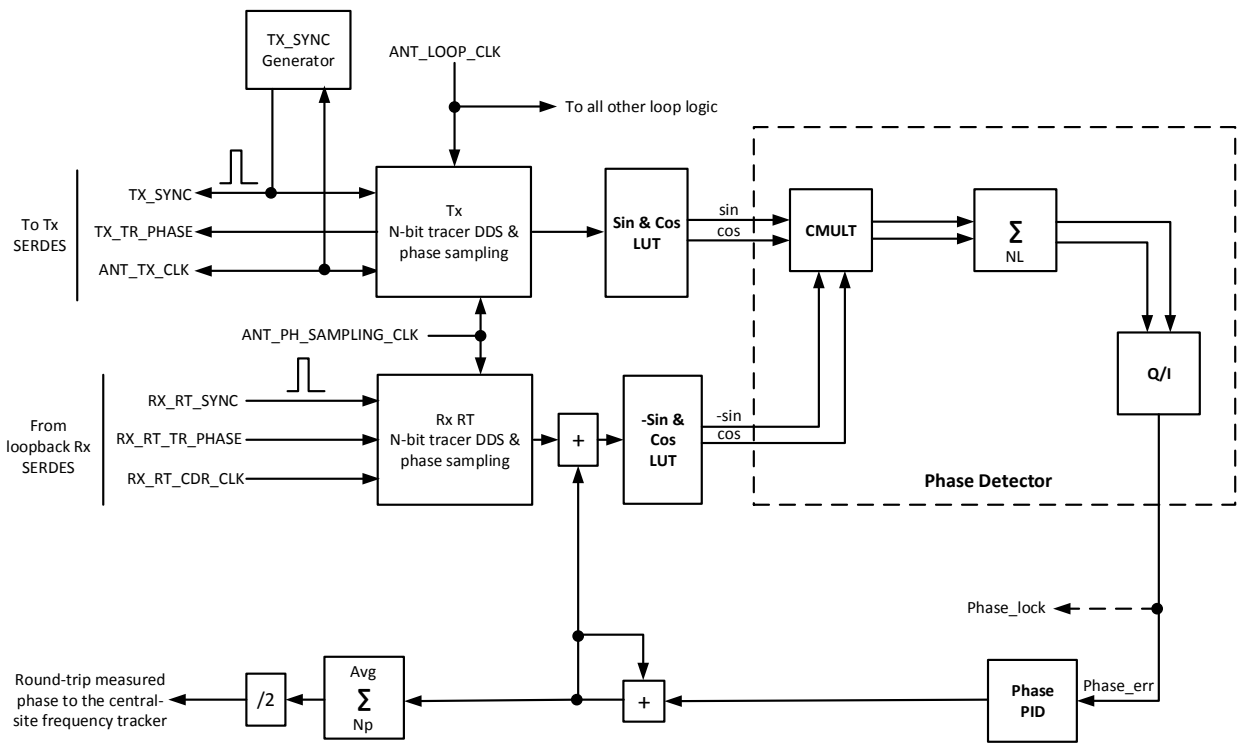

*Figure 4-8  Round-trip phase tracker block diagram.*

ANT_PH_SAMPLING_CLK and ANT_LOOP_CLK need not be derived from the aLO and can instead use an independent relatively low-quality LO since tracer phase is Nyquist zone-1 sampled.  However, what is not shown in Figure 4-8 is that the ANT_LOOP_CLK[14] is used for the round-trip delay measurement, essentially counting the time between the TX_SYNC pulse, and the return copy RX_RT_SYNC.  By counting in the aLO clock domain, this measurement can be corrected at the central site, using the measured aLO frequency, if required.

---

[14] Or the higher frequency ANT_PH_SAMPLING_CLK if higher time resolution is required.

The above block diagram also shows the starting point for tracer phase generation and the method for synchronizing all tracer DDSs. The "Tx N-bit tracer DDS" has an arbitrary initial phase and is free running in the antenna clock domain (ANT_TX_CLK). The TX_SYNC generator generates a pulse on a periodic basis, which captures the current tracer DDS phase TX_TR_PHASE. This is then transmitted to the central site using an "IC_telem" protocol such as that shown in Figure 4-20. There, the tracer phase is loaded into the tracer DDS of the Frequency Tracker. This same IC_telem stream is looped-back at the central-site and is loaded into the "Rx RT N-bit tracer DDS" at the antenna. In this manner, the "genesis" tracer DDS at the antenna, the tracer DDS at the central site, and the RT tracer DDS all remain synchronized to ensure continuity of phase and frequency measurements. If link loss occurs, signaling is lost but is re-established once the link comes back up. However, of course, during link loss the central Frequency Tracker and round-trip tracer DDS has no Rx CDR PLL clock and so it is no longer tracking. Notions of how to seamlessly fail-over a short term link loss are discussed later.

A discrete-time MathCad model of the round-trip phase measurement circuit of Figure 4-8 was run for a ~10 MHz tracer phase variation of ~7e-5 cycles (pk) at 2.5 Hz (i.e. $f_{c\_LPFF}$ for Tau_c~=0.1 sec as previously discussed), which needs to be tracked and that is ultimately fully compensated by injection as a cancelling term in the Frequency Tracker's Feedback DDS phase offset input. To this was added ~1 nsec pk-pk total (uncompensated) link jitter at ~100 Hz and ~2.2 kHz, which are ultimately rejected by the LPFF. A frequency-domain low-pass filter[15] with $f_c$~=3 Hz and -80 dB reject-band is used to perform the net result of LPFF filtering (assuming no Frequency Tracker rejection, the worst-case.) Phase-loop PID coefficients are set to $2^{-17}$, $2^{-29}$, and $2^{-23}$. 7e-5 cycles of phase at 2.5 Hz corresponds to an ADEV of ~5e-11 at Tau=0.1 sec for 300 km of aerial fiber [9]. Results of running this model are shown in the following figures:

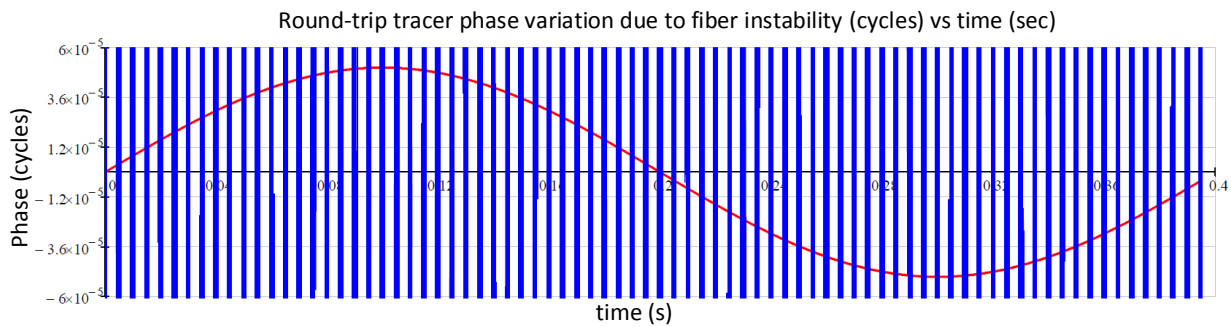

*Figure 4-9 Modeled tracer phase variation due to fiber instability. The **red** trace is the variation that must be tracked, if not it will be confused with aLO variation. The **blue** trace is variation that must ultimately be rejected in the post-Frequency Tracker LPFF output; its total variation is ~100X the vertical scale shown.*

---

[15] i.e. FFT of the entire simulation data set, multiply by the low-pass filter function, then inverse FFT.

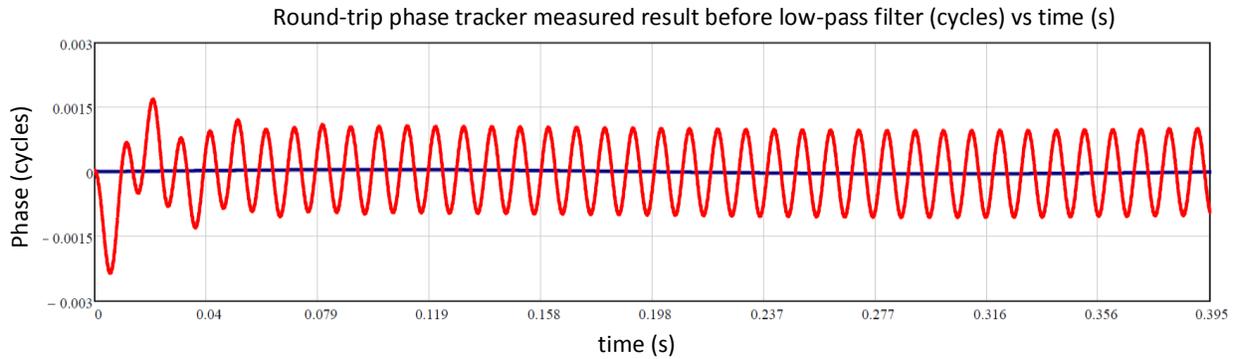

*Figure 4-10  Round-trip phase tracker output phase vs time (**red** trace) and model of the 2.5 Hz variation that must be tracked (**blue** trace.)  Much of the unwanted link jitter (at 100 Hz and 2.2 kHz) has been rejected by the loop, however much remains amounting to ~3e-4 cycles of RMS phase error, over 1000X more than allowed.*

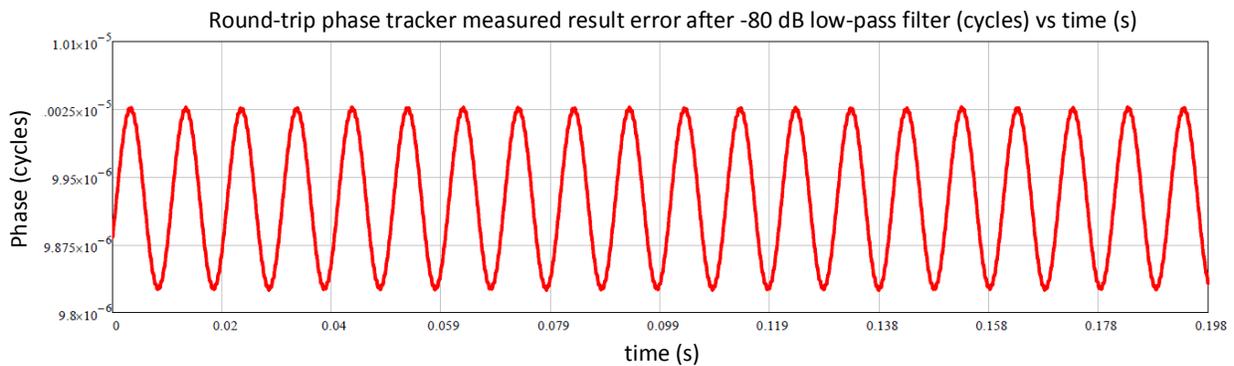

*Figure 4-11  Error in round-trip measured phase after -80 dB low-pass filter (cycles) vs time.  A phase offset of ~1e-5 cycles is present due to the frequency domain low-pass filter, however such is irrelevant. The RMS phase error (i.e. σ) is 7e-8 cycles (@ 10 MHz), which corresponds to ~2.5 deg at 1 THz.  Clearly the slow 2.5 Hz variation is being adequately tracked even in the presence of far higher link jitter.  The time scale on this plot is ½ of Figure 4-10 since only the latter ½ of the data is filtered and analyzed to eliminate confusion from loop initialization/stabilization.*

If needed, more LPFF stop-band attenuation can be employed but at some increased delay through the filter, knowing that the ReS FIFO (Figure 4-17) delay can always[16] be increased to match the LPFF delay, (i.e. fm=1 in equations (1) and (2)) thus not affecting aLO stability requirements.  For example, a -100 dB LPFF is easily possible, given the 18b+ multipliers available in modern FPGAs, and has a ~15% increase in delay and number of taps compared to a -80 dB filter[17].  This filter, normally multi-stage decimating, is light on FPGA resources since raw frequency measurements are input every ~10 usec.

To see phase tracking of the variation that must be measured due to a changing fiber delay and to see phase noise due only to discrete-time phase sampling and digital processing, the following plots are presented:

---

[16] Well, of course, there is a limit, however DDR4 memory has enormous capacity of 256 GB.

[17] 0.1 bandpass, 0.001 dB passband ripple, ~2.5% transition band, using the free on-line "TFilter" FIR filter design program.

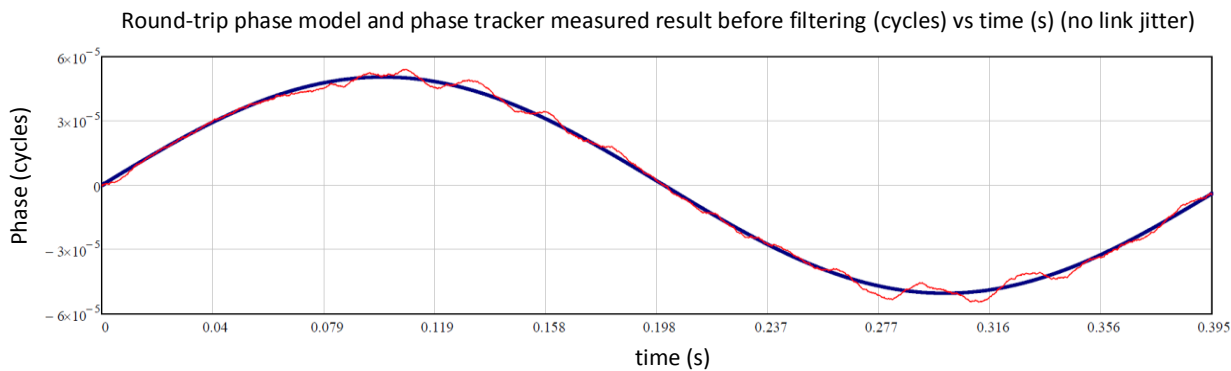

*Figure 4-12  Phase model and raw phase tracker measurement before low-pass filtering with no imposed link jitter.  This shows tracking performance and phase noise due to digital phase sampling and processing.  Without filtering the RMS (i.e. σ) phase error is ~1e-6 cycles (@~10 MHz.)*

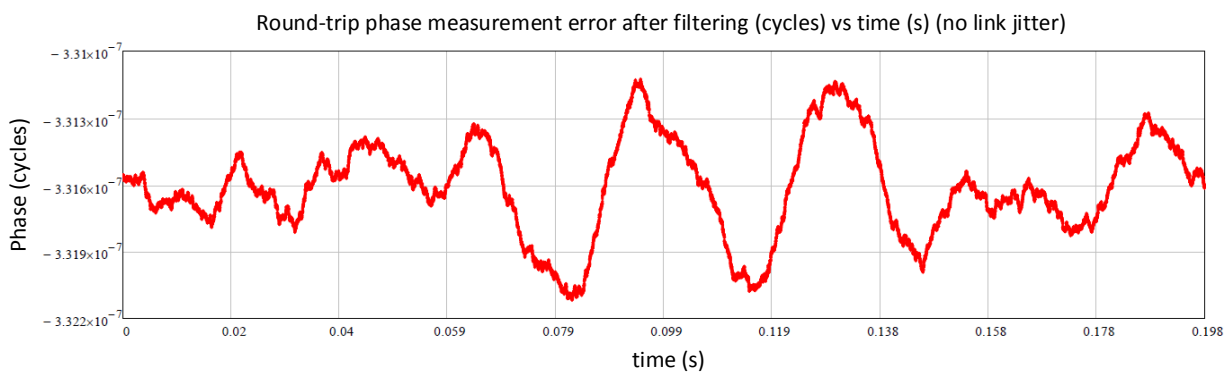

*Figure 4-13  Phase tracker-measured error, cycles vs time, after -80 dB low-pass filtering, no unwanted link jitter.  The RMS (i.e. σ) phase error is ~2e-10 cycles (@~10 MHz.)  The time scale on this plot is ½ of Figure 4-12 since only the latter ½ of the data is analyzed after filtering to eliminate apparent errors from loop initialization/stabilization.*

## 4.3  Tracer DDS and discrete-time phase sampling

In the Frequency Tracker block diagram of Figure 4-2 the tracer phase, from the tracer DDS, must be sampled into the reference clock domain (the REF_PH_SAMPLING_CLK and its decimated & phase-locked companion, REF_LOOP_CLK.)  This is a critical operation and is quasi-analogue in nature since tracer phase is modulated by aLO variations as well as fiber link-induced perturbations.  Both of these effects must be adequately captured, for any aLO offset from the reference frequency[18], including momentarily very low or no offsets.  Clearly, these latter two occurrences, although they should be rare, can occur and cause beating as first reported in [10], which was the case early-on in IC research effort.

Since that report, a superior tracer DDS and phase sampling architecture has been developed that shows no hint of beating effects during discrete-time simulation, for no, very low, some, and large frequency offsets from the reference frequency.  A case in point is Figure 4-3, where the offset from the ~10 MHz tracer frequency is ~2.8 Hz and clearly no beating occurs.

---

[18] Within the frequency tolerance of the antenna LO, typically +/- a few ppm.

In digital logic it is regularly necessary to transfer data between independent clock domains. There are generally two methods to do so: 1) dual-port memory and, 2) single-bit asynchronous sampling. In 1), the memory cell that contains the data to be transferred stores the information and digital logic on either side accesses it without contention from the other side except that, of course, the memory cell can't be in the process of being changed by one side whilst the other is trying to determine its state. In 2) a single register bit, which can be 0 or 1, is sampled by a register clocked in the other clock domain. If the register bit is changing state as it is being sampled in the other clock domain, that sampling may be successful (i.e. capture the change) or not, but it is accepted and expected that either case may occur. To ensure this transaction takes place it is required that the bit is 0 or 1 long enough to eventually be captured. This method can and does induce a metastability in the capturing register but this is quickly resolved due to the physics of the logic implementation as well as with additional register stages, 2 shown in Figure 4-14, although up to 5 may be more appropriate depending on the launch and capture relative clock rates and their clock rates relative to the clocking registers' gain and speed. Even if the rare clock-domain-crossing error occurs, it will be a glitch that will a) be filtered-out by the LPFF and b) possibly cause a momentary de-coherence for a period of time vastly shorter than the correlator integration time. An Internet search reveals many articles on the nature and behaviour of metastability—what is clear is that it is well understood and managed with proper digital logic design methods and that in no case does it result in a register bit being "stuck" in a state forever[19]. Also, it should be clear that it is fundamentally impossible for there to be a net cumulative phase bias in the sampled phase, meaning that it is impossible for there to be a phase drift (i.e. frequency offset) in the sampled phase that is not in the tracer DDS output.

An extension of method 2) is to transfer N bits of data at the same time, but with the restriction that only one of the N bits can change state on any transmit clock cycle and that the receive clock cycle must be sufficiently shorter than it to ensure no bit transitions are missed. This is a well-known method referred to as "Gray-code", employed where appropriate and with timing requirements described for example, in [11].

Discrete-time phase sampling employed uses both method 1) and 2) in concert to allow any number of phase bits, with each phase state any arbitrary value, to securely and reliably cross asynchronous clock domains. This greatly increases the parameter space for tracer frequency, number of phase bits, and performance. A block diagram of this method, including the tracer DDS, is shown in Figure 4-14 and based on the above description it should be clear how it works:

---

[19] In the literature, but also the author's personal experience in the design and function of the JVLA WIDAR correlator, where transferring bits across asynchronous clock domains was extensively used in FPGAs and the correlator ASIC. In that system there are millions if not billions of these cross-clock domain transactions reliably occurring every second and have been for the last ~10 years.

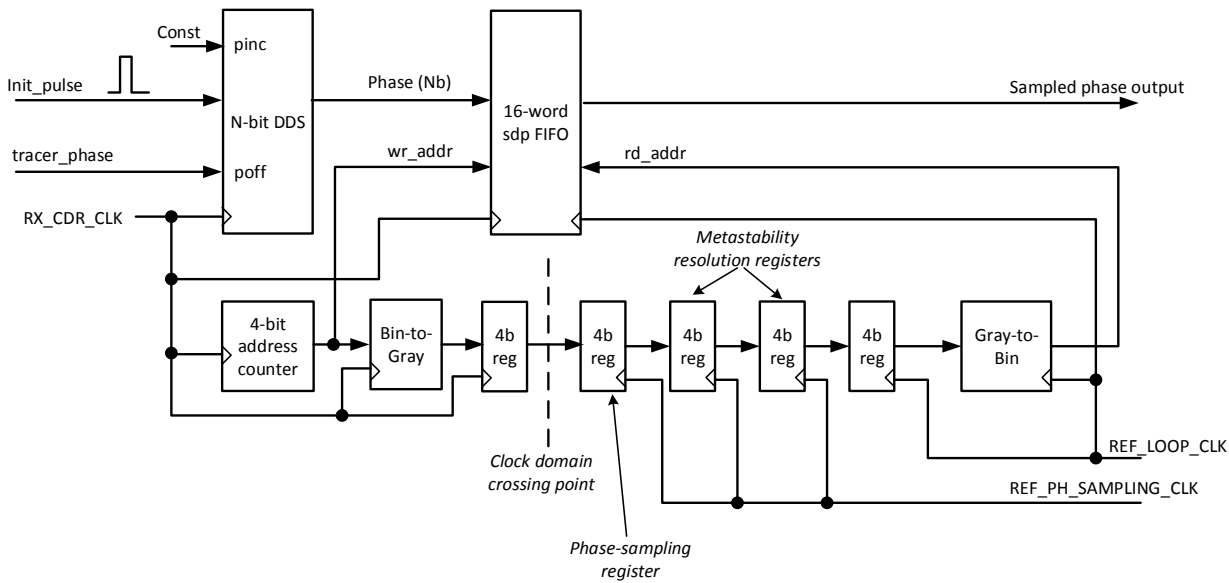

*Figure 4-14  Tracer phase sampling block diagram, including the tracer DDS and simplified IC signaling to it.  This circuit can be further simplified by consolidating the 4-bit address counter and the Bin-to-Gray blocks into a single Gray-code counter on the transmit side and using the sampled Gray-code for the rd_addr on the receive side.*

Critical high-performance timing is limited to a few, in this example[20], 4-bit registers and these can operate up to ~700-800 MHz (Intel Stratix-10) and even higher in an ASIC.  These registers must operate at the full clock rate without parallel/demux operations; the tracer DDS may be pipeline or multi-phased (e.g. "ping-pong") for higher net clock rate operation.  The "sdp FIFO" (simple dual-port) may be structured and operated similarly.

An early "first LED test" of the Frequency Tracker, complete with discrete time phase sampling, was performed using the Xilinx ZCU111 RFSoC evaluation board being used for the lab demo reported in [3].  This test used the Frequency Tracker to measure one of the board's crystal frequencies in the reference clock domain of another of the board's crystals, indicating key circuit status by driving board LEDs.  This was a quick test to empirically determine if the Frequency Tracker and phase sampling were working, although no data was acquired since such acquisition requires software, which was not available[21].  A block diagram of the FPGA design for this test is shown in Figure 4-15:

---

[20] And used in the laboratory demonstrator [3] FPGA code.

[21] As of this writing much software has been developed; the RFSoC FPGA design is complete, tested, and undergoing the usual iterative compile/modify process to achieve timing closure.

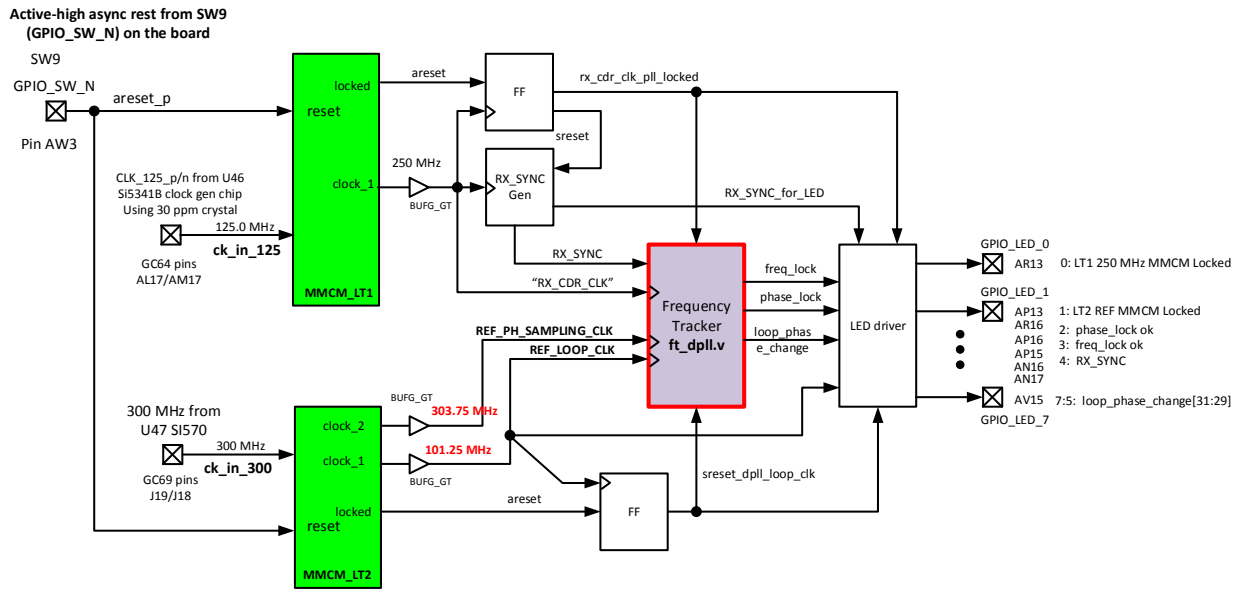

*Figure 4-15  Block diagram of the "first LED test" run on the Xilinx ZCU111 evaluation board RFSoC FPGA.*

The first LED test was successful, with LEDs lighting up as expected quickly after circuit start up. Whilst this doesn't quantify the performance of the Frequency Tracker, phase and frequency lock, determined internally in the circuit by analyzing internal variables to produce lock indicators (shown as "Phase_lock" and "Freq_lock" in Figure 4-2), would simply not have been asserted if lock were not established.

The phase stability of the REF_PH_SAMPLING_CLK at the *Clock domain crossing point* and for the additional *Metastability resolution registers*, relative to all other such instances in the same or other devices across a complete radio astronomy system, matters—if there is low-level differential phase wander that is within the passband of the LPFF, it will show up as antenna-based phase wander in correlator output visibilities. Thus, REF_PH_SAMPLING_CLK distribution across the system must be handled sufficiently well to ensure these effects are sufficiently low. The RX_CDR_CLK's dynamic phase behaviour also matters—changes in it that are within the passband of the LPFF, that are not due to the aLO, will also show up in correlated visibilities as above. Thus, the performance the of round-trip phase measurement to remove fiber delay perturbations, the value of the parameter "rtm" in equations (1) and (2), the phase noise performance of the SERDES Rx CDR PLL, and the phase noise of reference clock distribution must be considered. These topics are discussed in some detail further on.

## 4.4   The ReSampler

All-digital re-sampling of a digital signal from one sample frequency to another uses the same digital integer-sample and fractional-sample delay methods used in radio astronomy correlator wavefront delay trackers. The re-sampler contains a FIFO, followed by a fractional delay interpolator, and it retards or advances the read pointer, in coordination with fractional delay interpolation, to keep the total end-to-end delay constant and, of course, to keep the read and write FIFO pointers from colliding. A graphic showing the equivalent DAC/ADC circuit and its fundamental operation is shown in Figure 4-16; one can immediately see that the delay through this circuit is constant since there are fundamentally no delay elements. All-digital operations avoid the uncertainties that would be introduced, not to mention performance limitations, cost, and power, if actual DAC and ADC devices were used.

National Research Council Canada

Conseil national de recherches Canada

Canada

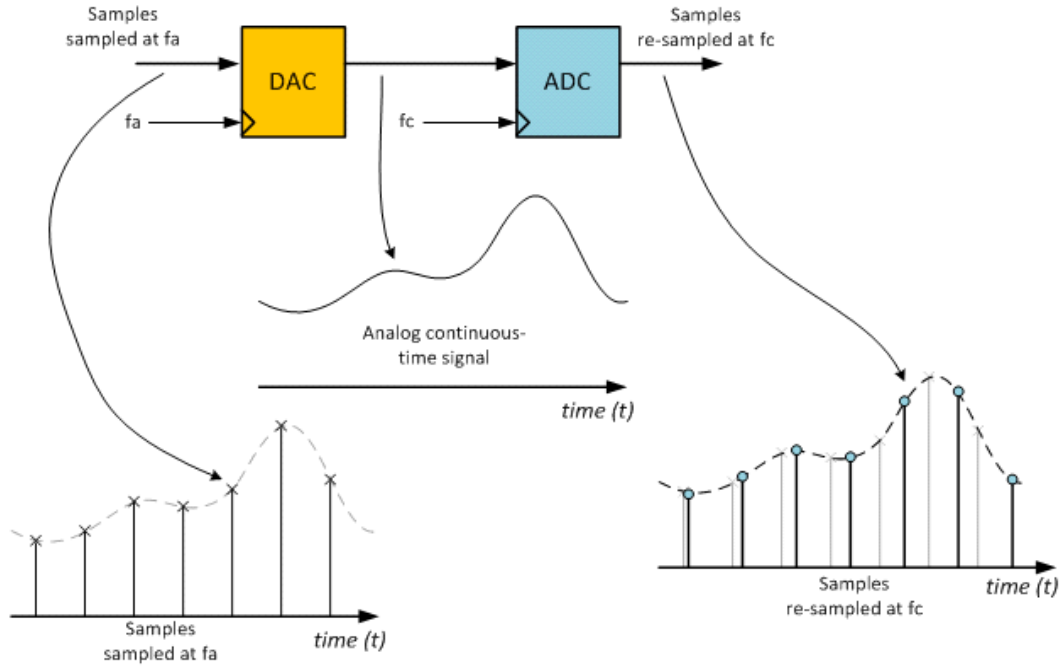

*Figure 4-16  Re-sampler DAC/ADC equivalent circuit and interpolation operation converting digitized science data from the fa clock domain to the fc clock domain.*

A simplified view of the "ReSampler", which includes the "ReS FIFO" that facilitates the read pointer advance/retard operation mentioned, fractional delay interpolator, and integrated Phase Rotator, is shown in the typical implementation block diagram of Figure 2-2.  A more detailed block diagram of the ReSampler is shown in Figure 4-17 and a description of its operation follows.

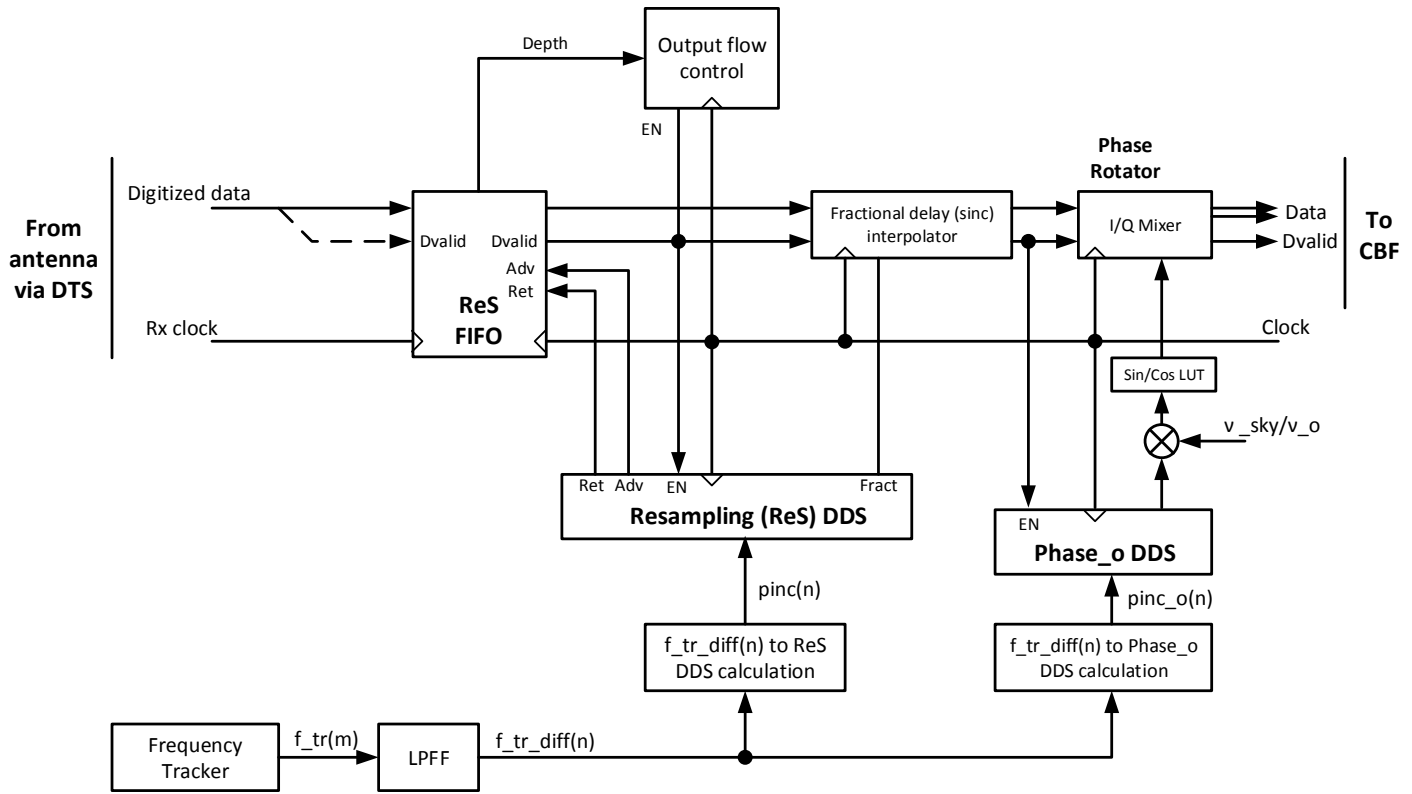

*Figure 4-17  Block diagram of the "ReSampler", which includes an integrated Phase Rotator, but doesn't include the Frequency Tracker or LPFF, shown here for context.*

The Frequency Tracker, shown in in Figure 4-2, produces a steady stream of high-time resolution tracer frequency measurements f_tr(m) into the LPFF.  In the lab demo (described in [3]) FPGA firmware, they are produced every ~10 μsec and are 64-bit fixed-point unsigned values with a 36-bit fractional component, providing ~1e-11 Hz of frequency resolution and contain no frequency bias relative to the reference clock.  As shown there, these measurements contain confusing fiber link jitter and, as shown in Figure 4-3 and Figure 4-4, contain phase noise from digital phase sampling of Figure 4-14.

Before the LPFF, 64-bit f_tr(m) numbers are trimmed to 43-bit signed "frequency difference from nominal" numbers (with the same 36-bit fractional component as above), "f_tr_diff."  Such is done for LPFF logic and processing efficiency, allowing for absolute frequency tolerance of the aLO, relative to the reference clock, of +/-10 ppm when using a ~10 MHz tracer frequency.  If needed, this tolerance could be expanded with more f_tr_diff bits.

After the LPFF, the filtered Frequency Tracker measurements, "f_tr_diff(n)" are fed to calculators that convert these to "pinc" values fed to the DDS that drive the ReSampler and the Phase_o Rotator, implemented as a digital I/Q mixer, yielding a complex output.  As shown, one calculator is for the Resampling DDS and one is for the Phase_o DDS.

The Phase_o DDS produces the phase correction for v_o, a conveniently-chosen frequency.  This phase is then multiplied up to the phase correction at the current v_sky, which is the (nominal) net down-conversion frequency of the signal being corrected.  For this to work, the Phase_o DDS output (phase) must contain the total phase, integer+fractional, since timing chain (see highlighted paragraph on p. 32) start up. This seems onerous, but with enough Phase_o DDS bits is easily implemented. For example, if v_o=1.0 GHz, f_Phase_o_DDS=125 MHz[22], and a +/-10 ppm aLO crystal is used, the worst-case phase change of Phase_o DDS output over 10 yrs is 1 GHz x 1e-5 x 125MHz x 3600 s x 24 hrs x 365 days x 10 yrs ~= 4e20 cycles. This requires ~68 bits to capture. Thus, with 68 bits of integer phase and 32 bits of fractional phase, the Phase_o DDS, for this example, requires a 100-bit DDS—not a difficult logic implementation at 125 MHz—but potentially requiring a timing chain restart every 10 years[23]. The v_sky/v_o multiplier then needs to be a 100-bit x Nb_ratio multiplier, where the ratio of v_sky/v_o contains an integer and fractional part. For v_sky_max = 1THz, with a 1 MHz v_sky tuning resolution, ~7 significant figures are required, likely meaning Nb_ratio~=16 bits is more than sufficient, TBC. Even if there is a low-level error in v_sky/v_o digital representation, this error multiplies phase, resulting in a small phase error with no cumulative phase drift component. As well, the 100-bit x 16-bit (in this example) multiplier will require several pipeline stages to yield a result, resulting in a time delay in application of the phase calculation to the I/Q mixer. This delay is constant in the reference clock domain, resulting in no net frequency-dependent effect. Finally, it should be clear that when v_sky changes to some other tuning and then comes back, no net phase jump occurs since the Phase_o DDS operation has been continuous, meaning that coherence on the sky has been maintained.

The DC gain of the LPFF must be precisely known and applied as a rescaling factor after the LPFF, to deal with bit growth that naturally occurs during LPFF (FIR filter) processing. If this rescaling isn't precise, it will cause the ReSampler output to drift in delay and phase. One method that works well in practise is to scale the tap coefficients such that the sum of the (integer) tap coefficients is precisely a power of 2; post-LPFF rescaling then simply involves a bit shift and unbiased rounding operation, guaranteeing required gain accuracy.

For the Resampling DDS, the pinc equation is (see Section 13 Appendix for derivation):

$$pinc_{ReS\_dds}(n) = \frac{2^{(nb_{ReS\_dds} + nb_{tr\_dds})} \cdot R_{a\_t} \cdot f\_tr\_diff(n)}{f_{ck\_ReS\_dds} \cdot pinc_{tr\_dds}} \qquad (5)$$

Where:

$nb_{ReS\_dds}$ is the number of bits in the ReSampler DDS.

$nb_{tr\_dds}$ is the number of bits in the tracer DDS.

$R_{a\_t}$ is the ratio of the ADC sampling frequency to the clocking frequency of the tracer DDS, where both have no error (i.e. their nominal frequency.) E.g. if the ADC is digitizing at 40.0 Gs/s in the aLO clock domain, and the tracer DDS at the central site is clocking at 625.0 MHz using the SERDES Rx CDR PLL-recovered clock, this ratio is 64. Whilst a power of 2 in this ratio is convenient, it is not essential.

$f\_tr\_diff(n)$ are the streaming measurements out of the LPFF. If the ADC frequency is higher than nominal, this will be a positive number, if lower, a negative number.

---

[22] i.e. "Clock"=125 MHz, although normally the DDSs operate at sub-Clock frequencies, since Clock is normally the highest frequency that the FPGA can support for the main sample-processing pipeline.
[23] Although likely never required if a +/-1 ppm aLO crystal is used and knowing that its frequency will be wandering around between these limits, rather than stuck at one extreme indefinitely.

$f_{ck\_ReS\_dds}$ is the clocking frequency of the ReSampler DDS.

$pinc_{tr\_dds}$ is the fixed pinc value chosen to be used in the tracer DDSs across the system.

**Example:** *for the lab demo described [3]: $nb_{ReS\_dds}$=64; $nb_{tr\_dds}$=16; $R_{a\_t}$=1; $f_{ReS\_dds}$=125.0e6; $pinc_{tr\_dds}$=2693; and f_tr_diff(n) measurements are in the range of +/-127 Hz, with a 36-bit fractional component as described above.*

The above calculations are done with fixed-point arithmetic and in any case must ensure there are no biases introduced into the resulting calculation. Biases will accumulate and cause a drift of the re-sampled output, which will show up in correlated visibilities and eventually cause the over/under-flow of the ReS FIFO if significant enough. Biases in this calculation, which should only happen during the division operation, shouldn't occur due to the dithering effect of a noisy f_tr_diff value as in Figure 4-5. Examination of the lab demo [3] end-to-end FPGA RTL simulation waveform data for both the pinc value fed to the ReS DDS of Figure 4-17, and the pinc value fed to the Feedback DDS of the Frequency tracker of Figure 4-2, indicate that all bits are being "tickled", shown in Figure 4-18 and Figure 4-19 below. This does not guarantee that there is no bias, but is an indication of such. Dividers for both of these pinc calculations are implemented using the Xilinx IP Catalog "Divider Generator", described in the Xilinx document "PG151." These descriptions, for the "divider solutions" used, do not seem to explicitly indicate if there are biases in the results or not, but if so should be, in the worst case, at the level of some fraction of a pinc LSBit. For example, the bias of one pinc LSBit of a 64-bit ReS DDS operating at $f_{ReS\_DDS}$= 250 MHz, after one year leads to a delay error of 4e-4 samples and after 10 years, 4e-3 samples. Similarly for the Phase DDS, but in cycles of phase. Nevertheless, further investigation into this issue is needed and such is identified as a risk in section 10.

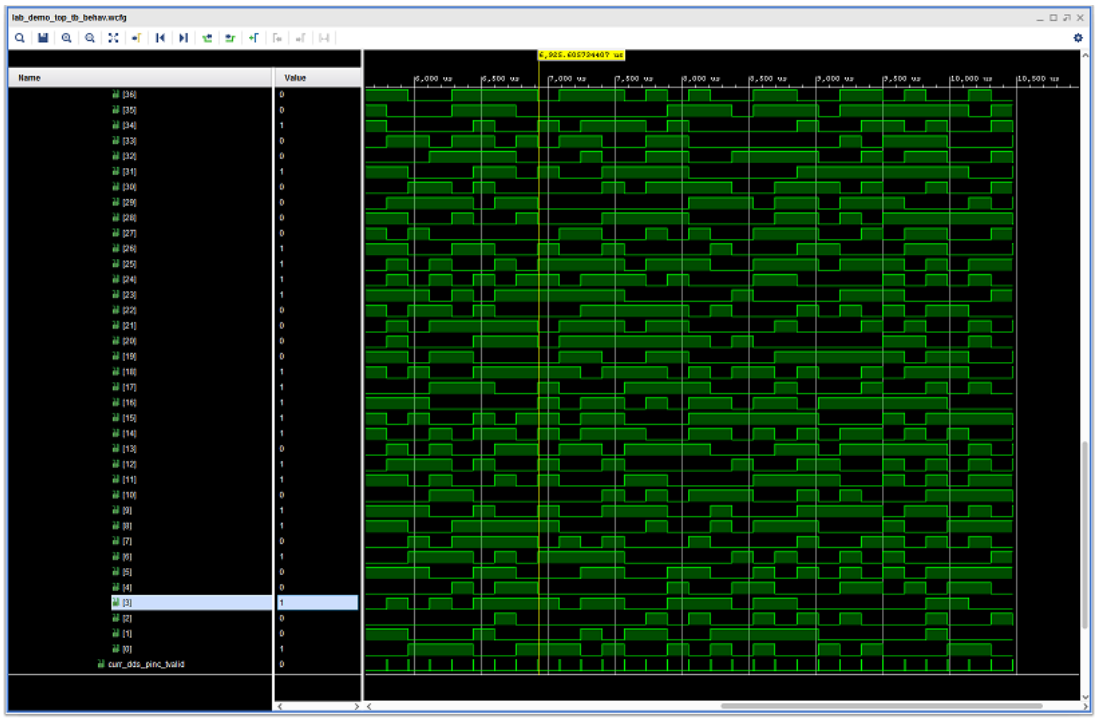

*Figure 4-18 FPGA RTL simulation waveforms for the lower 37 bits (of 64 bits) of pinc fed to the ReS DDS indicating that all bits are being "tickled" due to the dithering effect of a noisy f_tr_diff on equation (5).*

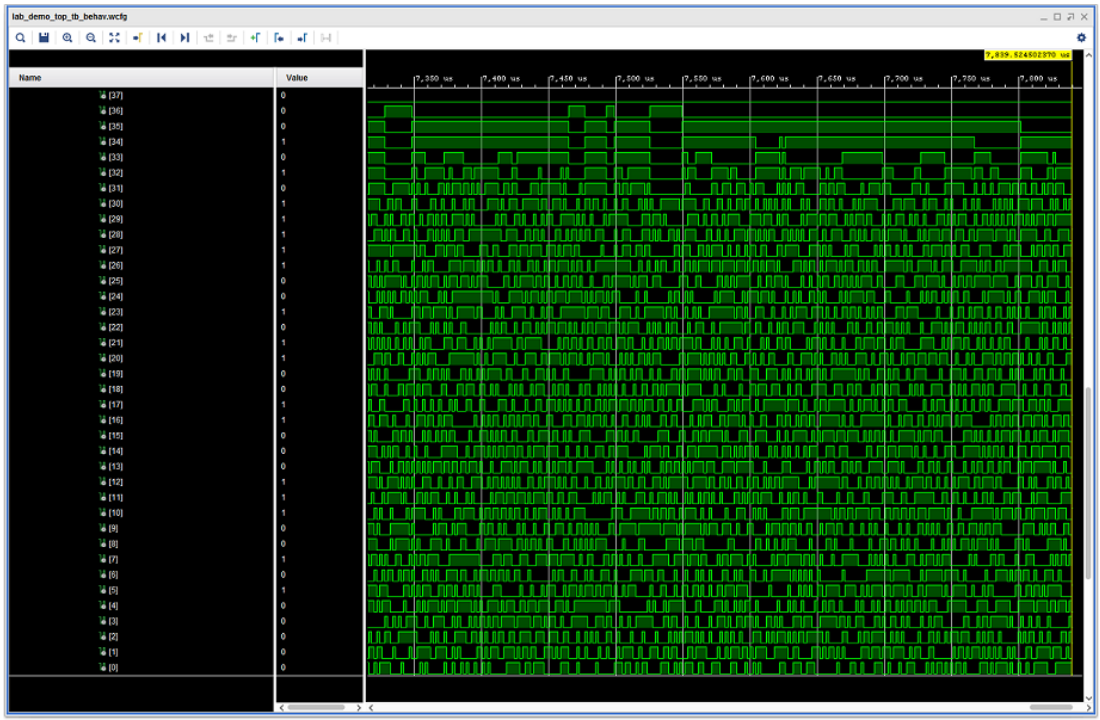

*Figure 4-19  FPGA RTL simulation waveforms for the lower 36 bits (of 64 bits) of pinc fed to the Feedback DDS of the Frequency Tracker (Figure 4-2) indicating that all bits are being "tickled" due to the dithering effect of a noisy f_tracer_loop.*

The Phase_o DDS $pinc_{ph\_DDS}$ calculation is similar except that $R_{a\_t}$ is instead $R_{v\_o\_t}$, where $R_{v\_o\_t}$ is the ratio of the conveniently-chosen v_o to the clocking frequency of the tracer DDS. $pinc_{ph\_DDS}$ is a positive number if the netLO at the antenna (i.e. v_sky) is higher than nominal and negative if it is lower than nominal. This calculation assumes that the digitized data has the same frequency sense as the sky; if not, a simple "flip the sign of every other sample" correction can be done at an appropriate place, easiest to understand if it is done before the Phase Rotator. For ALMA there is an added complication and that is 180 deg and 90 deg Walsh function switching at the antenna. These phase jumps are in the data and flow through the pipeline to the output of the Phase Rotator seamlessly, requiring no consideration in this circuit.

For both DDSs, their output phase vs time functions reflect the difference between the actual antenna frequency, which is changing with time, and the nominal, which is in the reference clock domain and is therefore constant. For example, if the antenna ADC clock frequency is precisely 1.0 Hz higher than nominal, no matter what the actual ADC clock frequency, the ReSampler DDS output will be a sawtooth function at precisely 1.0 Hz, as measured in the reference clock domain. Similarly for the Phase_o DDS, multiplied up by v_sky/v_o.

Referring to Figure 4-17, in order for the ReS FIFO to not overflow or underflow, the write data rate and the read data rate must be matched on average. The ReS FIFO is written, and the write pointer advanced, for every sample[24] arriving from the antenna. This is accomplished by adjusting the read pointer to "keep up with it" by advancing (Adv) or retarding (Ret) the read pointer, compared to where it would've been without any such adjustment. Such Adv/Ret pulses derive from the Resampling DDS whenever the paccum phase ramp rolls over from max to 0 or vice versa. When the DDS pinc is a positive number, meaning that the antenna ADC frequency is higher than nominal and it is writing into the ReS FIFO faster than nominal, Adv pulses occur; when negative, Ret pulses occurs. The "Output flow control" circuit simply ensures that the ReS FIFO doesn't over or underflow by asserting Dvalid low every so often, going to all of the blocks shown. Thus, the actual "Clock" frequency used for this pipeline in the FPGA must be > the highest data rate that can be written into the FIFO. If desired, the Output flow control circuit could be a DDS, generating Dvalid lows at the required rate. For example, a Clock frequency of 501 MHz, for a data rate of 500 Mwords/s could be used, easily meeting a 10 ppm aLO tolerance requirement[25].

---

[24] Or "super-sample rate" group of samples.

ReSampler output discrete-time samples then enter the "Fractional delay (sinc) interpolator", where the fractional delay (ramp) from the ReS DDS selects the set of tap coefficients required to apply the correct interpolated value. When Adv or Ret pulses happen, the fractional delay flips accordingly such that there is no net delay change across the transition. This operation, when it happens, can[26] be made entirely glitchless [12][13][14] such that, save for the fidelity of the fractional delay coefficients, there is no discontinuity in the data.

> *The reader might observe that this is not a stateless circuit and such is indeed the case—the digital signal chain from the ADC sampling point in the antenna until the output of the ReSampler is part of the timing chain of each antenna[27]. If there is an undetected glitch or discontinuity in this chain, then coherence on the sky is lost. Various methods are available and should be used to "bullet-proof" this circuit: PLLs' lock indicators; data transport sample counting and error detection; detection of unusual discontinuities in Frequency Tracker f_tr(m) measurements; comparison of a counter value in and out of the ReS FIFO to detect any slips, etc. The robustness of this chain and the IC measurement timing chain, including momentary link dropouts, is discussed later.*

As mentioned, the fractional part of the Resampling DDS is used to drive the interpolator, an all-pass FIR filter (a.k.a. "sinc interpolator") whose output are interpolated samples between antenna ADC samples. Typically this filter has 1024 steps [14] between +/-0.5 samples or 0 and 1 samples of delay since FPGA resources are adequate. However, for this application, minimizing FPGA resources is important[28] because of the high sample rate, and therefore a more prudent approach must be taken. Generally, for discrete-delay-step interpolation the phase error is a sawtooth function with sensitivity loss of (1-sinc($\phi_{pp}$))[29] [15]; for the upper edge of the re-sampled bandwidth, this is (1-sinc(2·$\pi$/Nsteps)). For Nsteps=16, the sensitivity loss is ~2.6% at the upper band edge (much lower when sinc function is integrated across the band), which should be acceptable to minimize hardware implementation cost. This doesn't take into account additional steps that might be needed to ensure dynamic range, for example for RFI.

---

[25] This also brings up the point that the pipeline from the ADC output to the ReSampler & Phase rotator output need not run at a clock frequency derived from the aLO since it is all discrete-time processing.

[26] The mechanics of which, though quite simple, are not shown in Figure 4-17.

[27] The other part being the IC measurement timing chain which includes link integrity and calculations, i.e. round-trip phase tracker in the antenna and Frequency Tracker at the central site.

[28] Although, see section 9.1 for a notion of how it might be integrated with the correlator 1st-stage filterbank.

[29] Where $\phi_{pp}$ must be the net baseline phase noise, having a uniform distribution.

For ALMA with ENOBs=5 [1], Nbits/sample=5 could be used, but allowing for another 1b per sample to Nbits/sample=6, an FPGA block memory structure can be used for the FIR interpolator, a well-known technique where the address is formed from the data (6 bits) and interpolation step (4 bits), and the output is the product of the two. For example, the Intel Stratix-10 FPGA has "M20K" memories that can be configured in various ways, including 1024 x 20-bit and 2048 x 10-bit. Thus, one such memory is required for each tap of the FIR, this allows for 6-bit data and 32-steps of interpolation with a 10-bit output (sufficient for ~60 dB of dynamic range), improving the sensitivity loss to 0.16%.

The number of taps required for the FIR is also a design parameter and depends on bandpass ripple, band-edge transition-band, bandwidth coverage, and the number of interpolation steps[30]. In the ReSampler described in [16], 64 taps are sufficient. For ALMA 2030, assuming a 40 Gs/s 5-ENOB ADC, 10-bit LUT output, 64 taps, clocking the FPGA FIR at 500 MHz (net rate, including Dvalid lows described above), the number of M20Ks required is: 40e9/500e6 x 64 x 2 = 10,240, for both polarizations. For the Stratix-10 SX2800, 14 nm FinFET c. 2015 technology, 11,721 M20Ks are available, clearly showing that the ReSampler FIR interpolator alone consumes almost the entire device's M20K block RAM. However, this is the bulk of the processing needed for the entire IC implementation.

The Sin/Cos LUT, which converts Phase DDS output to the required complex vector for the Phase Rotator should, even for coarsely-quantized data, be many bits to ensure precision of application even for cases where the aLO frequency is very close to, or at, the nominal frequency (i.e. $f\_tr\_diff$~=0.) In modern FPGAs, filled with large 18b+ multipliers, this LUT should take full advantage and be at least 16b.

Both the ReSampler fractional part and Sin/Cos LUT outputs are, in any case, finite precision. This means that there will be a saw-tooth error function between the required function and what is actually implemented in finite-bit logic. This can cause spectral splatter of narrowband signals in the digitized data and a method to handle it ([17][13][14]), works very nicely, even for strong RFI. This method entails randomizing (scrambling) the least-significant bit (LSB) of the interpolator fractional value and the Phase DDS output before the Sin/Cos LUT and essentially turns the saw-tooth function and spectral splatter into white noise. Provided the pseudo-random function feeding the scrambler is different for each antenna and doesn't repeat for a very long time, this noise does not correlate.

Finally, for ALMA the input to output pipeline of Figure 4-17, for a 40 Gs/s ADC, is "super-sample rate" (SSR), i.e. it consists of lanes of time-demultiplexed consecutive-time samples, clocked at a frequency the FPGA can operate at. Appropriate parallel-processing methods must be used, touched on above. This may include operating the Resampling DDS and/or Phase DDS in multi-phase fashion, needed when their DDS output value changes appreciably across a lane of samples.

For example, a 40 Gs/s ADC processed at a 500 MHz FPGA clock frequency has 80 lanes. With an aLO tolerance of +/-10 ppm, the phase ramp present at the ReSampler DDS could be as high as 400 kHz. Thus, the sample interpolation change across 80 lanes is 400 kHz x 80/40e9 ~= 0.0008 samples which is << an interpolation step size of $1/2^{(ENOB+2)}$ ~= 0.0078 samples for ENOB=5, obviously not requiring SSR operation of the DDS. The ReSampler DDS clock frequency also needs to be high enough so that there isn't an appreciable change in the fractional delay between its clocking. For example, if the ReSampler

---

[30] i.e. at some point the number of interpolation steps is the limiting factor and adding more taps to reduce ripple has no net benefit.

DDS is clocked at 250 MHz, it means 400 kHz/250MHz = 0.0016 of a sample worst-case change per 250 MHz clock cycle, which is sufficiently less than the 0.0078 sample interpolation step size.

For the Phase DDS, with ALMA observing at 1 THz with an aLO tolerance of +/-10 ppm, its output phase ramp is at a maximum frequency of 10 MHz, which is a 10 MHz x 80/40e9 ~= 7.2 deg phase slope across the 80 lanes. Thus, an SSR phase output should be used, where each phase is 1/80 fraction of the DDS pinc added to the DDS phase output, although 80 different phases is likely not needed.

## 4.5    IC tracer synchronization and signaling—"IC_telem" protocol

As mentioned previously, tracer DDSs are synchronized at both ends of the link, and round-trip measured phase is transmitted to the central site on that same fiber link. A simple serial "IC_telem" protocol that carries such signaling is shown in Figure 4-20, used in the lab demo discussed in [3]. Payload bits in this frame are scrambled and error-checked with a CRC-16 code, ensuring data integrity over any communications link.

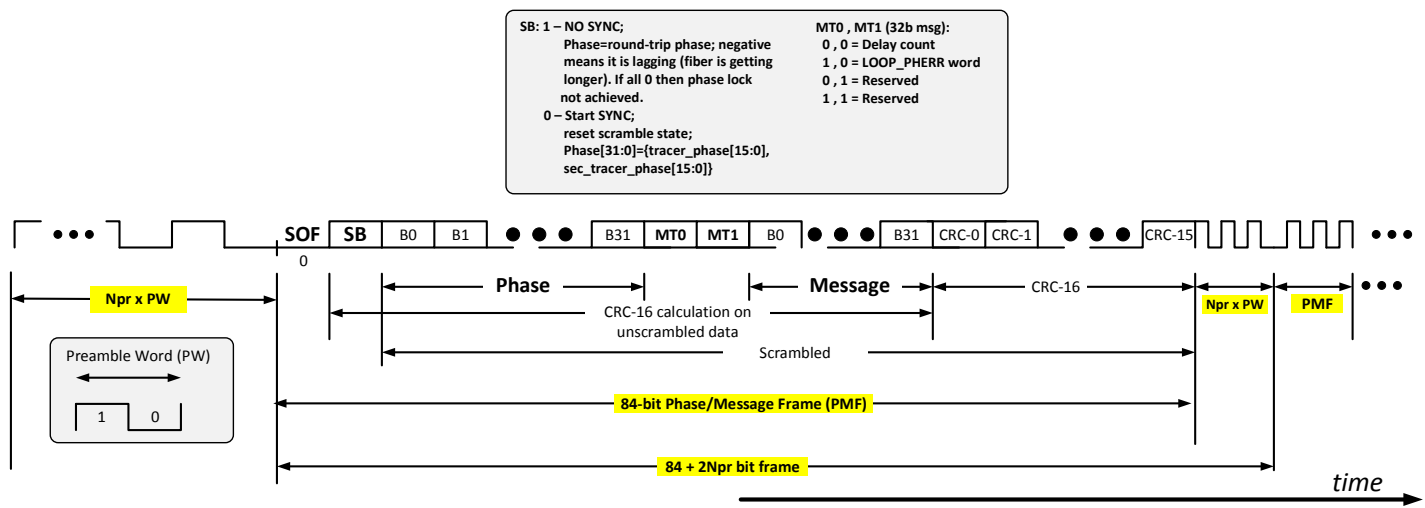

*Figure 4-20 "IC_telem" protocol used in the lab demo discussed in [3]. This protocol is used for reliable IC signaling between the antenna and the central site. Minimum Npr is probably ~4, but may be adjusted to meet system frame cadence requirements, if any.*

The "Start SYNC" (SB=0) bit indicates that the 32-bit Phase message contains the tracer phase, shown in Figure 4-2, Figure 4-8, and Figure 4-14, loaded into the DDSs to keep them synchronized. The Phase message allows for 2, 16-bit tracer phases, the primary for the DDSs operating at ~10 MHz, and a secondary operating at a much lower frequency to facilitate momentary link loss fail-over, as discussed later. Otherwise when SB=1, the Phase message contains round-trip measured phase, streaming at a rate determined by the link serial data rate, "Npr" (in the above figure), and the occupancy of the serial stream if integrated with digitized science data. As mentioned previously, for the lab demo discussed in [3] frame cadence is ~10 µsec, providing ~100,000 round-trip phase measurements per second to the Frequency Tracker.

The 32-bit Message field content is determined by the MT bits and is primarily used to transmit the round-trip-measured delay to the central site for, among other things, initializing the "Static fiber link delay compensation" in the Frequency Tracker of Figure 4-2. This delay measurement may also be used to effectively provide the antenna with a time-epoch marker, by either transmitting a pre-negatively-delayed marker to the antenna[31] (as an IC_telem Message), or by delaying the time-epoch marker at the central site by an appropriate amount before insertion into the data stream for downstream correlator/beamformer synchronization and timing; the choice of which is determined by IC architecture, discussed later.

The 32-bit Message field can, of course, be used to convey other user information across the link. Shown in Figure 4-20 is "LOOP_PHERR", which is the captured streaming "Phase err" of Figure 4-8, providing one indicator of the quality of the RT phase measurement[32].

# 5   Dynamic end-to-end behaviour

It is informative to develop an intuitive understanding of the dynamic end-to-end behaviour of data flow from the ADC output in the antenna until the ReSampler output. What follows is a qualitative discussion of this behaviour for the typical case of Figure 2-2 (which is the same as "ALMA Architecture A") wherein the digitized science data is clocked, end-to-end, at a frequency locked to the aLO. This behaviour also applies with some modification to other possible IC architectures since the principles are the same.

Once the end-to-end pipeline from the ADC output to the ReSampler output is active and tracking, the absolute delay in time between these two endpoints is constant and invariant to changes in aLO frequency and fiber delay, to within the required interpolation accuracy of the ReSampler. This is clearly illustrated in the correlator output of the lab demo discussed in [3] and, of course, must be the case otherwise any delay variation will show up as unwanted residual delay behaviour in correlator output visibilities.

A simplified view of the end-to-end pipeline is shown in Figure 5-1 (a). This includes fixed register (i.e. "D flip-flop") delays at both ends, the fiber, the ReSampler FIFO, and of course the interpolator (ReSampler fractional delay.) Fixed registers at either end, on the face of it, would seem to have a delay that changes with aLO frequency changes and therefore the delay through them is not fixed. This is true of course, but if fixed registers are only used then no re-sampling to another sample rate is possible, and the pipeline is effectively broken. It is also true that the sample rate out of the fixed registers is identical to its input and therefore, for example, the behaviour of the ADC output is exactly reflected at the output of the fixed register delay after it and so no ADC behaviour is lost. What seems to be clear is that fixed registers are equivalent to memory cells and, to simplify the end-to-end model to that of Figure 5-1 (b), the fixed registers are subsumed into, and become part of, the input of the ReS FIFO.

---

[31] i.e. so it shows up at the antenna coincident with the required time epoch, e.g. UTC 1-second.

[32] i.e. when the RT phase servo is active and tracking, Phase_err should show no noticeable biases or oscillations on timescales passed by the LPFF. If it does, it means that there are measurement errors in the Frequency Tracker LPFF output that are not intrinsically in the antenna LO —generally meaning that the RT phase servo PID gains are not high enough.

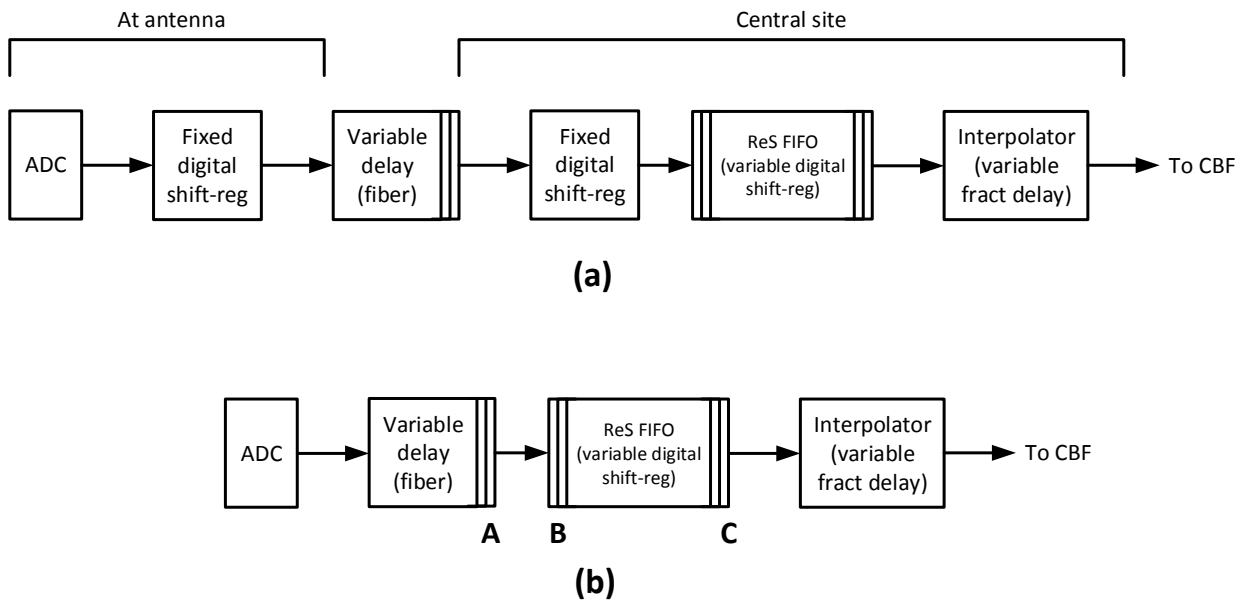

*Figure 5-1  Dynamic end-to-end behaviour diagram.  The "Fixed digital shift-reg" blocks in (a), are effectively absorbed into the input section of the ReS FIFO, resulting in the simplified view shown in (b).*

Considering Figure 5-1 (b), the interaction between the variable delay of the fiber manifesting itself at point A, and the input of the ReS FIFO at point B:

- The fiber delay is independent of the data rate on it.
- When the fiber delay changes, the frequency of the Rx CDR PLL-recovered clock at point B, used to clock the input of the ReS FIFO (not shown in Figure 5-1, but shown in Figure 4-17), changes and this causes an exact opposite compensating effect in the depth of the FIFO since the FIFO write address pointer changes from where it would've otherwise been.  For example, if the fiber delay increases, the CDR-recovered clock frequency drops, causing the write pointer to be retarded from where it otherwise would've been.  This means that the distance between the read pointer (which hasn't changed from where it would've been) and the write pointer has dropped and therefore the delay through the FIFO has been reduced by an amount corresponding to the increased fiber delay.
- This behaviour is further understood if the fiber and the ReS FIFO are place in a black box with the input to the black box at the ADC (output) and the output of the black box at point C clocked at the same frequency.  In this case it is clear that there is no insight into what is happening in the black box and therefore no difference between its behaviour and a direct hard-wired connection.

Considering Figure 5-1 (b), the behaviour of the ReS FIFO, with different instantaneous input and output clock rates (points B and C) (and without considering the variable fiber delay discussed above):

- The absolute delay through the FIFO in units of time is constant as long samples are read from it at the same rate as they were written.  For example, if samples are being written at a constant rate f_write and read at the same constant rate, i.e. f_read=f_write, then clearly the delay is constant.  Now with the FIFO operating in that manner, start writing to the FIFO at 2 x f_write— twice as many samples are being written per unit time, but the time-occupancy of each one is now one-half of previous, and so the total delay through the FIFO has not changed. On the read side, when the 2 x f_write samples appear, as long as they are read-out at the f_read=2 x f_write, the same time-occupancy of samples is being removed from the FIFO as before and therefore its delay has not changed.

- When f_write changes, the depth of the FIFO *in samples* will increase and decrease accordingly.
- The delay through the FIFO in units of time can be determined by the sum of all sample times, of all samples in it. If f_read≈f_write for the depth of the FIFO, this calculation can be approximated to be f_read/depth, something that can be calculated at any instant in time and with accuracy proportional to the difference between f_read and f_write and the depth of the FIFO. Since f_write is not known, and an accurate measure of (essentially) f_read is available at the output of the LPFF where it is known, the calculation is done with f_read.

> *This means that **the total delay from the ADC to the ReSampler output can be calculated, at any time, as long as the pipeline is active and stable** if the fiber delay, # of fixed digital registers, and instantaneous ReS FIFO depth, as well as f_read are known. This delay is constant and the accuracy of the calculation—and therefore jitter of successive calculations—depends on the depth of the FIFO, how long it takes to acquire that depth, how much f_write changes over the delay through the FIFO, the accuracy of the f_read measurement (derived from the LPFF output), and, of course the change in fiber delay during the acquisition time.*

- As long as the net f_read rate matches the net f_write rate, the FIFO will not overflow or underflow. What this means is that the rate at which the FIFO is read can be governed by a controller ("Output flow control" in Figure 4-17) monitoring its depth and issuing ENables at an appropriate rate, to ensure it doesn't over or underflow. In interactions with the interpolator and the ReS FIFO and Phase DDSs, each valid output sample from the interpolator is at the re-sampled rate and "Clock" in Figure 4-17 can be from any convenient source as long as its frequency is higher than the maximum possible rate out of the FIFO.

> *Thus, the entire circuit of Figure 4-17, except for the Frequency Tracker, can be located at any convenient place in the entire ADC output to ReSampler output timing chain, including at the antenna.*

# 6    Central-site timing distribution to Frequency Trackers

The phase stability of the REF_PH_SAMPLING_CLK at the *Clock domain crossing point* of Figure 4-14, relative to all similar locations across the entire telescope, matters. Any differential that is not rejected by the LPFF will show up in correlator output visibilities as antenna-based delay/phase wander. If high enough, this will cause unwanted de-coherence and/or phase wander that cannot be calibrated-out. Thus, distribution of the "Central common LO" to Frequency Trackers across a system, shown in Figure 2-2, is important for correct operation.

There are two basic methods possible for reference clock distribution from the Central Timing Reference (CTR) to the Frequency Trackers. The first is simply via passive one-way cables, all approximately matched in length. Provided that the temperature of all of the cables is kept approximately the same, they all have the same CTE (coefficient of thermal expansion), and depending on allowed differential phase wander, this method is the desired method since it is the simplest. However, for a large system where cables (fiber) might be quite long and are too expensive or impossible to thermally match, a better method is needed.

This better (second) method is active round-trip measured clock distribution, shown in Figure 6-1:

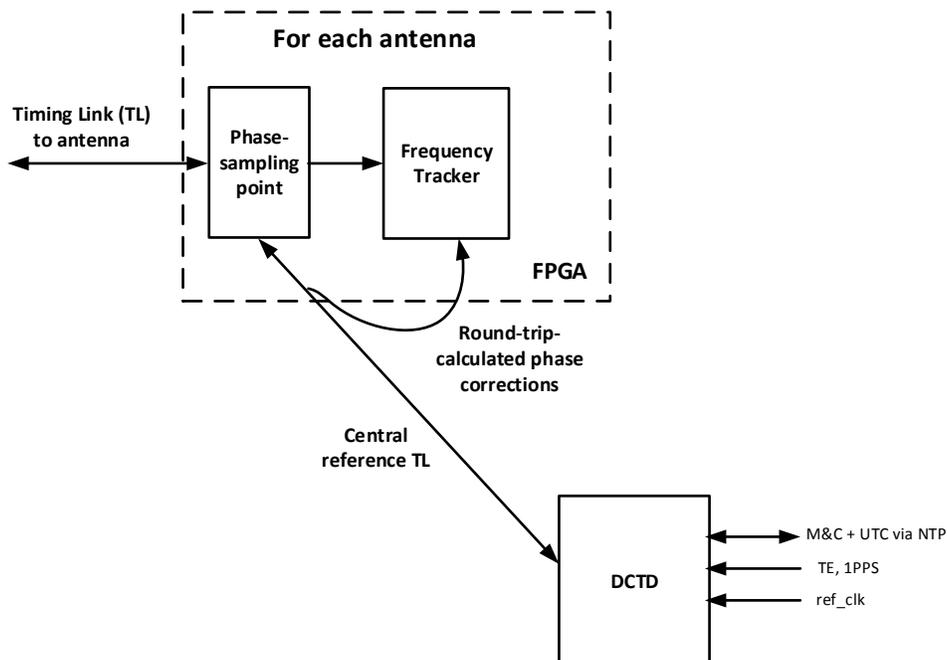

*Figure 6-1 Active round-trip measured method. Round-trip measured phase, across each full-duplex distribution cable independently, is measured and sent to the Frequency Tracker as a phase correction into the Feedback DDS of Figure 4-2, in the same manner as the round-trip measured phase corrections from the antenna.*

In this method the clock and round-trip-measured phase is transmitted to each phase sampling point from a central distributor—Digital Central Timing Distributor (DCTD)—using a similar or identical protocol as shown in Figure 4-20. There, the CDR PLL-recovered clock is used (directly or via a frequency synthesizer) as the REF_PH_SAMPLING_CLK (and REF_LOOP_CLK) clock shown in Figure 4-2 and Figure 4-14. REF_PH_SAMPLING_CLK is wandering around in phase, of course, and the round-trip-measurement at the DCTD, for each phase sampling point independently, is fed into the Feedback DDS of the associated Frequency Tracker (Figure 4-2), along with the round-trip measured phase from each antenna. This works as follows:

- The sampled-phase phase vs time ramp, coming from the (antenna) tracer DDS, and the phase ramp out of the Feedback DDS are lined up/matched when the loop is active and tracking.
- A phase change of REF_PH_SAMPLING clock will therefore sample a different phase in the tracer DDS output than it would have had there not been a phase change. For example, if the phase of REF_PH_SAMPLING_CLK lags by a bit, it advances the sampled phase since it samples it at a later time.

- A phase change of REF_LOOP_CLK, which is phase-locked to REF_PH_SAMPLING_CLK, will change the phase of the phase ramp out of the Feedback DDS. For example, if the phase of REF_LOOP_CLK lags by a bit, it retards the sampled phase since the DDS phase has been incremented less.
- Both the REF_PH_SAMPLING_CLK and the REF_LOOP_CLK, since they are phase-locked, will therefore proportionally change the phase difference between the sampled tracer phase ramp and the Feedback DDS phase ramp, and this change will be identical to the round-trip measured phase from the DCTD, i.e. twice the REF_PH_SAMPLING clock at the end point.
- Thus, the round-trip measured phase from the DCTD, when applied as an offset to the Feedback DDS phase, will keep the sampled phase ramp and the Feedback DDS phase ramp aligned and invariant to the changing delay of the DCTD-to-phase sampling and Feedback DDS fiber.
- The frequency of the DCTD-source clock and therefore its derived tracer, will be tied to the reference clock frequency. Also, if the circuit of Figure 8-3 is required, it will be at a higher frequency than the (aLO-driven) tracer DDS. This means that DCTD round-trip-measured tracer phase is not the same as the compensating phase that needs to be applied. However, this difference can be easily converted to the correct phase with a "similar triangles" calculation using the output f_tr_diff from the LPFF and the actual REF_PH_SAMPLING frequency appropriately.
- After this correction, the wandering phase of the REF_LOOP_CLK doesn't matter since all Frequency Tracker loop operations are discrete-time and not sensitive to REF_LOOP_CLK phase wander. i.e. the Frequency Tracker loop operates in real time, but such operation is not sensitive to some phase wander in REF_LOOP_CLK, as in the above points just discussed.

> *The net result is that this method allows the Frequency Trackers to be located at convenient locations across virtually any size system. This degree of freedom may be a significant factor in selecting an IC architecture that is best for any radio telescope including, of course, ALMA.*

A simplified layout diagram of a DCTD, which could be used for ALMA, is shown in Figure 6-2. It is possible that circuitry could fit in a 1U 19" rack-mount box, although both front and rear panels would be needed to accommodate all 80 (antenna) SFP/SFP+ cages. A 2U box would allow all 80 SFP/SFP+ cages to be accessed from the front-panel, although a stacked/mezzanine solution would probably be needed in this case.

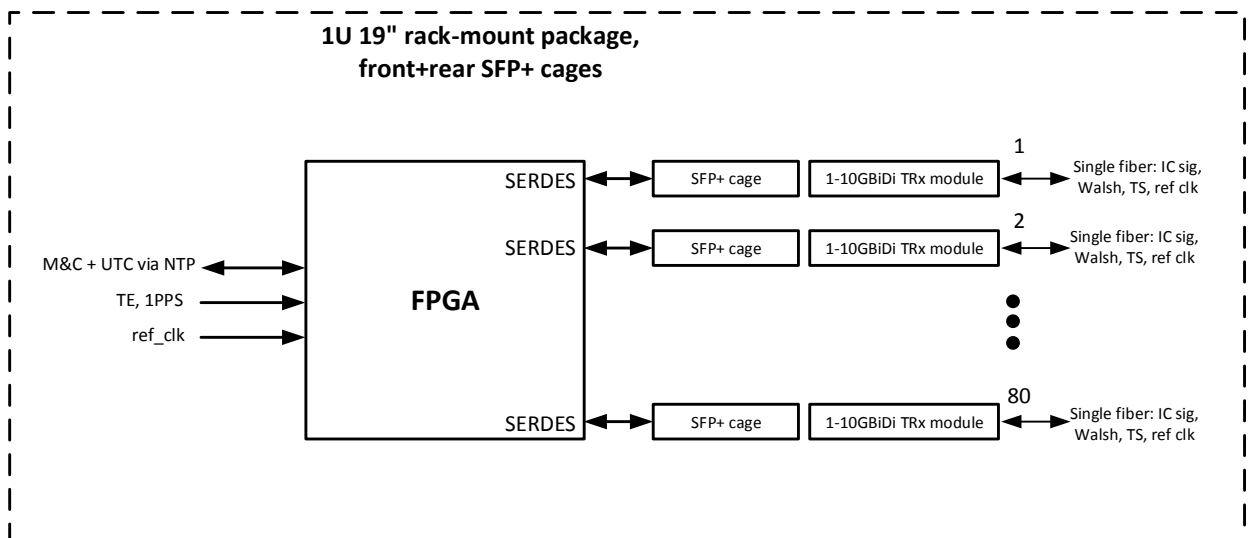

*Figure 6-2  Simplified layout diagram of a Digital Central Timing Distributor (DCTD) for ALMA.  All round-trip-measured phase calculations for 80 antennas can be done in one FPGA, with connections to 80 SFP/SFP+ modules, for BiDi single-fiber connection to each Frequency Tracker end-point FPGA.  1-10Gbps rates are shown, but a lower rate such as OC-12 622 Mbps (perhaps 625 Mbps since it is probably more convenient) or some other frequency, could be used.*

# 7   IC and DTS link-loss fail over operation

As mentioned, the timing chain of each antenna, as far as IC and the digital part goes, is from the ADC sampling point in the antenna until the output of the ReSampler at the central site.  Any interruption of this chain will result in loss of coherence of the antenna.  It would be beneficial from a reliability/robustness standpoint, to be able to allow for a momentary antenna-to-central site IC timing and data link loss and recovery without losing coherence.  A functional block diagram of how this might be done in digital logic is shown in Figure 7-1.  In this diagram:

- A LOS (Loss of Signal) detector determines if any one of the Rx CDR PLL lock, IC signaling, or streaming data from the antenna has been lost.  If so, it raises the tracer_LOS and/or data_LOS flags so that appropriate action can be taken.
- If only the data_LOS flag is raised, the "Local packet gen" keeps advancing the ReS FIFO write pointer, whilst counting the number of increments that occur, and reconciling these counts with the packet counter coming from the DTS once DTS packet streaming resumes.  This keeps the digital streaming data pipeline loaded without any discontinuities although, of course, there will be no useful data in the pipeline.
- If the tracer_LOS flag is raised (as well, or instead), then:
  - The Primary and Secondary Frequency Trackers flywheel at their previously known frequency, which means that their Feedback DDS pinc inputs (Figure 4-2) are frozen.
  - A drift will therefore start to occur between the frozen Feedback DDSs' phase outputs and the aLO, i.e. compared to what they would've been w/o an LOS.
  - During this drift, the Secondary Frequency Tracker tracer phase can only drift by up to ~+/-45 deg from where it would've been so that its Phase Detector output stays well within the right-hand-side of the complex plane.  To meet a defined maximum LOS time, its tracer frequency is chosen to be sufficiently low for the aLO ADEV chosen.  For example, for an ADEV with a +1 log-log slope starting at 1e-13 and Tau=0.1 sec, and an allowed link-loss time of 1000 seconds, the secondary tracer frequency needs to be <~40 MHz, for $\Delta\phi_{pk}\leq45$ deg (3$\sigma$.) Clearly, just the primary tracer frequency of ~10 MHz should be sufficient for ALMA!

o Streaming packets to keep the ReSampler pipeline loaded are as described above. It may be the case that they are coming in via the DTS faster or slower than the fly-wheeling that occurs and so logic needs to handle this case and seamless recovery involves reconciling DTS packet counters with ReS FIFO write address increments appropriately.

o Once the tracer_LOS condition goes away, the Secondary Frequency Tracker's Phase_err signal is appropriately scaled and drives (slews, without discontinuity) the Primary's loop, until the Primary's Phase Detector output is within +/-45 deg of the +'ve real axis at which time it uses its own Phase_err signal to close phase and start tracking the aLO changes.

This operation will cause potentially many cycles of primary tracer phase to be slewed, all of which affects (slews) the f_tr_diff(n) output of the LPFF that drives the ReS DDS in the ReSampler, slewing samples through the ReSampler pipeline accordingly. This is a seamless, continuous-time-like and necessary operation to get the primary tracer phase and the timing chain back to where it would've been had there not been a link loss.

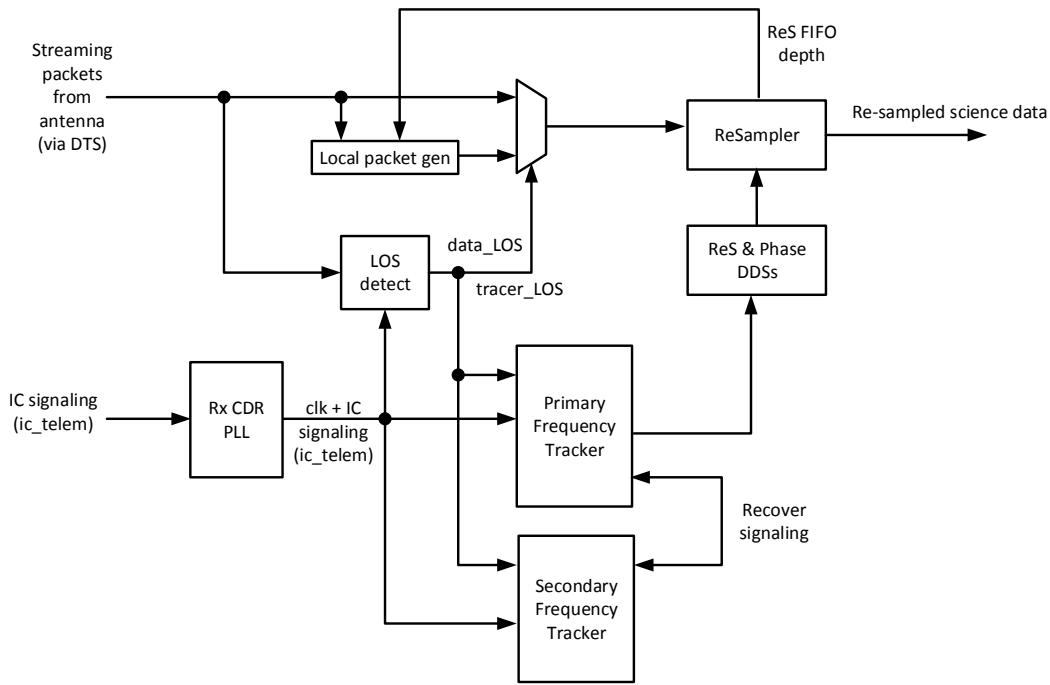

*Figure 7-1 Functional block diagram of IC timing and/or DTS link loss recovery digital circuit.*

This circuit has not been functionally tested, however its theory of operation and result—that of keeping the ReSampler data pipeline continuously loaded and Primary Frequency Tracker output measurements continuous in time without any uncertainties or discontinuities—seems clear.

In the vein of link loss recovery ability is also the potential ability for the entire aLO and timing chain to tolerate a momentary antenna power loss.  Details of how this could be achieved are well beyond the scope of this discussion except to say that doing so involves local battery backup of the LO and all frequency synthesizers and signaling that could introduce phase discontinuities or ambiguities in the timing chain if they undergo a power cycle.

# 8   FPGA Rx CDR PLLs and PLL frequency synthesizers

FPGA clocking circuits (Rx CDR PLLs, PLL frequency synthesizers for general logic and SERDES use), used but not explicitly identified in previous circuit discussions, are high gain and sufficient for digital purposes.  Rx CDR PLLs are designed to track significant jitter, specified in units of bits (UI—Unit Interval), up to very high frequencies, often 10 kHz and higher.   For example, the 10 Gbps jitter tolerance mask shown in Figure 8-1 indicates the minimum performance requirement of Rx CDR PLLs that recover a 10 GHz clock from a 10 Gbps serial stream and [18] is a test report from Xilinx (requiring special permission from Xilinx to access) for its FPGA SERDES Rx CDR PLL performance, exceeding that shown in Figure 8-1.  A similar plot for the Micrel SY87701 CDR PLL chip, which might be needed if the FPGA SERDES channel is not used or appropriate, is shown in Figure 8-2.

FPGA PLL clock frequency synthesizers, in the settings dialogs for them in the FPGA manufacturer's toolkit, can typically be set for high or low gain, frequency synthesis, or phase-locking to the input, balancing jitter with performance.  There is no ADEV or "phase noise offset from carrier" data for these, instead the data sheets or settings dialogs state input jitter tolerance and total pk-pk output jitter.

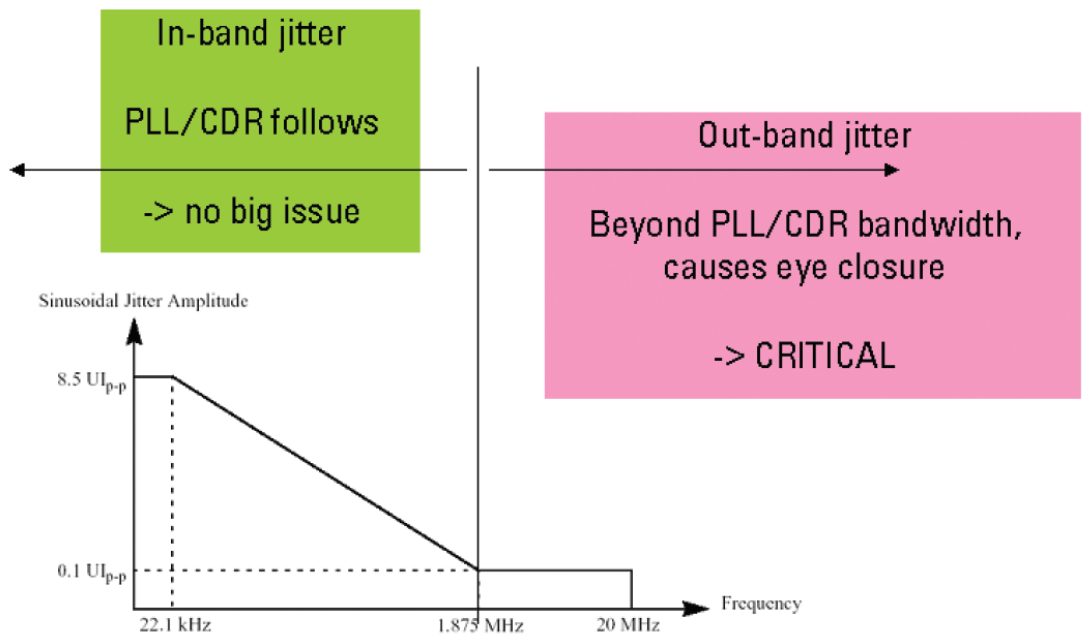

*Figure 8-1  10G jitter tolerance mask, according to IEEE 802.3ae XAUI, extracted from [19].*

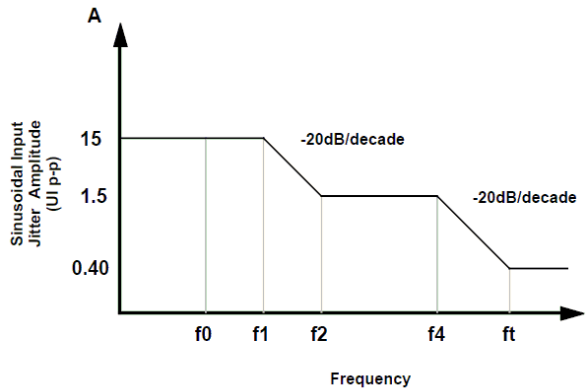

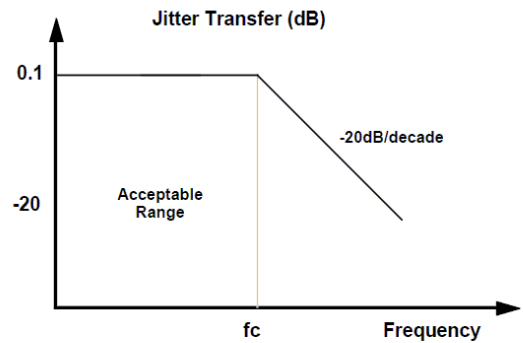

| OC/STS-N Level | f0 (Hz) | f1 (Hz) | f2 (Hz) | f3 (kHz) | ft (kHz) |
|---|---|---|---|---|---|
| 3 | 10 | 30 | 300 | 6.5 | 65 |
| 12 | 10 | 30 | 300 | 25 | 250 |

| OC/STS-N Level | fc (kHz) | P (dB) |
|---|---|---|
| 3 | 130 | 0.1 |
| 12 | 225 | 0.1 |

**Figure 1. Input Jitter Tolerance**            **Figure 2. Jitter Transfer**

*Figure 8-2 Jitter tolerance mask and jitter transfer function for the Micrel SY87701, extracted from the Micrel data sheet. This chip may be used if OC-12 (~622 Mbps) is used for IC signaling independent of the DTS. The Jitter Transfer plot indicates that all jitter up to fc~=225 kHz is passed (with some amplification) to the output, suggesting that this chip would be a good candidate for a 3R OEO repeater for longer reach, if needed.*

The statistics of FPGA PLL additive phase jitter (i.e. jitter relative to the input clock or serial stream) is important. As long as jitter is white/broadband in nature, then its total effect on phase noise in the output of the LPFF will decrease with decreasing $f_{c\_LPFF}$, e.g. Figure 4-3-Figure 4-5[33]. However, if there are discrete-frequency spurs in the PLLs' phase noise vs frequency offset from carrier that are ultimately within the LPFF passband, these spurs—i.e. narrowband phase oscillations that don't proportionally decrease with decreasing $f_{c\_LPFF}$—will cause confusion with aLO behaviour and will be a performance limitation if significant enough.

The Micrel SY87701 CDR PLL chip's data sheet contains no phase noise specifications (offset from carrier in dBc/Hz), just jitter tolerance as above. Nor does Intel Stratix-10 or Xilinx Ultrascale+ (DS-923) FPGA data sheets; direct attempts to obtain such information from the manufacturers have not yielded any information (notwithstanding their applications' engineers attempts.) The Silicon Labs Si5381/82 jitter cleaner device data sheet lists impressive phase noise for a ~122 MHz carrier down to 100 Hz offset from it, as well as total phase noise due to spurs from 1 MHz to 30 MHz offset. The Si5381/82 is a chip that might be used for providing an FPGA with its "cleaned" SERDES REF_CLOCK and uses a crystal oscillator as its reference to do so. Whether this device has spurs that lie within $f_{c\_LPFF}$ offset from carrier (translated to the tracer frequency) or not can likely only be determined with empirical testing. If unwanted spurs are present, then this chip can't be used for the stated purpose above, rather the clean FPGA SERDES REF_CLOCK should source from the aLO device as directly as possible.

---

[33] Indeed in the MathCad discrete-time models that produced these plots (and in FPGA RTL code simulations), jitter of this nature, and according to the manufacturer's specifications, was simply added as random cycle-to-cycle jitter. As mentioned however, phase noise due to discrete-time phase sampling is far higher.

Thus, it would seem that it is not possible to retire this risk via analysis alone. This risk can be retired or realized during actual hardware testing, but having one or more fallback positions that retires this risk, should it be realized, is important.

One possibility is to develop a custom CDR PLL out of discrete components without elements that can cause spurs, designing and testing it so that spurs within the LPFF passband are not present. Another possibility is shown in Figure 8-3, wherein no CDR PLL or PLL is used at all in critical TL paths. With the correct choice of IC signaling (e.g. IC_telem of Figure 4-20), transport frequencies, and separate but coupled "raw" clocks, it is clear that all PLLs in the critical timing paths have been eliminated. For example, 622 Mbps OC-12, could be used for a 622.08/2 ~= 311 MHz tracer DDS clock and 1.250 Gbps (1 GbE line rate), could be used as a 1250/2 = 625 MHz REF_PH_SAMPLING_CLK. Small frequency deltas from these, which will be immaterial to the transparent/non-re-timing optics involved, could be used if necessary. The Rx CDR PLL shown in the figure could be implemented external to the FPGA in a device such as the aforementioned Micrel SY87701 but its output clock frequency must be within the limits of the FPGA clock pin input, which is typically ~650-700 MHz[34].

The drawback from this fallback position is that it is likely not possible to seamlessly recover from a momentary link loss as discussed in section 7, and the number of possible IC implementation architectures for ALMA, discussed in the next section, is likely reduced to just Architecture D or E.

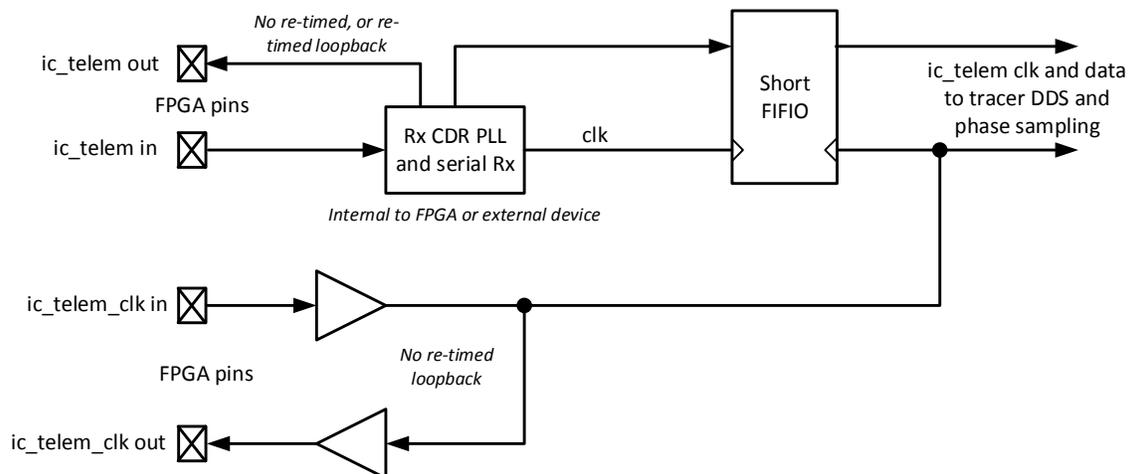

*Figure 8-3  Circuit in front of phase sampling in the Frequency Tracker that eliminates any FPGA PLLs or frequency synthesizers in the path to the Clock domain crossing point of Figure 4-14. A circuit, similar to this, can also be used for the REF_PH_SAMPLING_CLK and decimated REF_LOOP_CLK, both ultimately sourcing from the DCTD.*

# 9   IC implementation architectures for ALMA

Thus far, the entire discussion has been referencing the "typical IC implementation" shown in Figure 2-2. However, this is only one possibility and what follows is a discussion of the operation and pros and cons of five possible architectures that could be used for ALMA.

---

[34] For 14/16 nm class FinFET technology.

## 9.1 Architecture A

This architecture is the same as the typical IC implementation, but includes some ALMA-specific details. A functional block diagram of Architecture A is shown in Figure 9-1.

In this architecture the IC_telem protocol is integrated with the digitized science data providing framing and error checking by proxy. Figure 9-2 is a simplified block diagram showing one way this may be done. This method, used in the lab demo described in [3] allows the full data rate of the link to be used for digitized science data with no overhead, sacrificing an insignificant sensitivity loss of the science data to do so:

- Every $N^{th}$ sample the LSBit of the digitized science data is replaced by the IC_telem protocol bits and thus flows clear-channel to the receiver at the other end. The value of "N" is chosen considering sensitivity loss and the required IC_telem data/frame rate. The latter is a consideration if the IC_telem frame also contains TE and Walsh phase switching information, which it probably would in this case.
- The FPGA SERDES gearbox is operated in raw-mode, meaning that whatever is present at the parallel transmit interface is what is transmitted serially, without modification. Thus, the user is responsible for transmit data scrambling, as well as de-scrambling and framing at the receive end.
- At the receive end, the IC_telem frame is searched in receiver-deserializer and $n^{th}$ bit phase space. Once frame-lock is achieved[35], the science data, including replacement IC_telem frame, is de-scrambled to yield the clear science data and resulting in every $n^{th}$ LSBit being scrambled. Since the LFSR scramble code generator is restarted periodically (on the SB-bit of the IC_telem frame), there is some possibility of correlation and so further no-reset/repeat scrambling of this bit should occur, different for each antenna, to guarantee no correlation.
- The IC_telem CRC-16 code provides robust error checking for the IC_telem frame itself, but also for the entire link and therefore the science data. This isn't "banking transaction quality" error checking, however it should be sufficient for digitized radio astronomy signals since errors, at even quite high rates such as 1e-4, have a negligible impact on data quality and sensitivity loss and in any case won't correlate if the mechanisms producing the errors are independent as they should be.

---

[35] Assured with the CRC-16 error checking in the IC_telem frame.

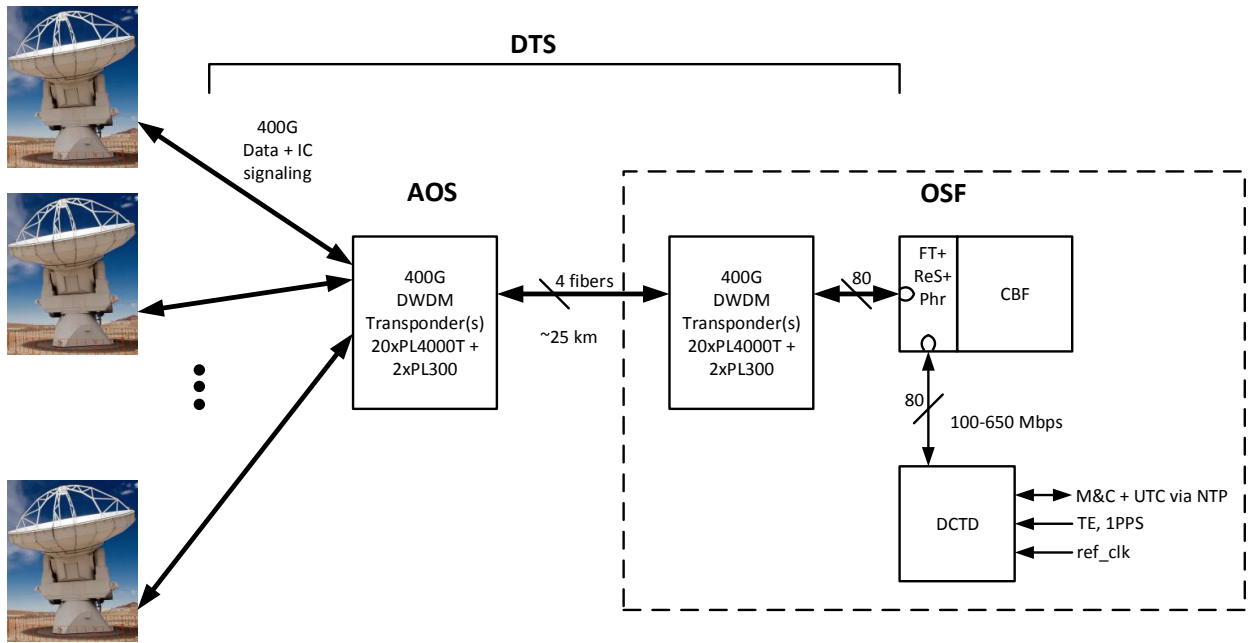

*Figure 9-1  Architecture A.  The IC_telem protocol is fully integrated with the digitized science data, with loopback locations indicated.  400G per antenna carries data for 40 Gs/s, dual-pol'n, 5 bits/sample. DWDM transponders provide the full-duplex link, for all 80 antennas, between the AOS and OSF using only 4 fibers.  COTS PacketLight transponders are shown—with more PL4000T equipment up to 128 antennas can be supported over the same 4 fibers.*

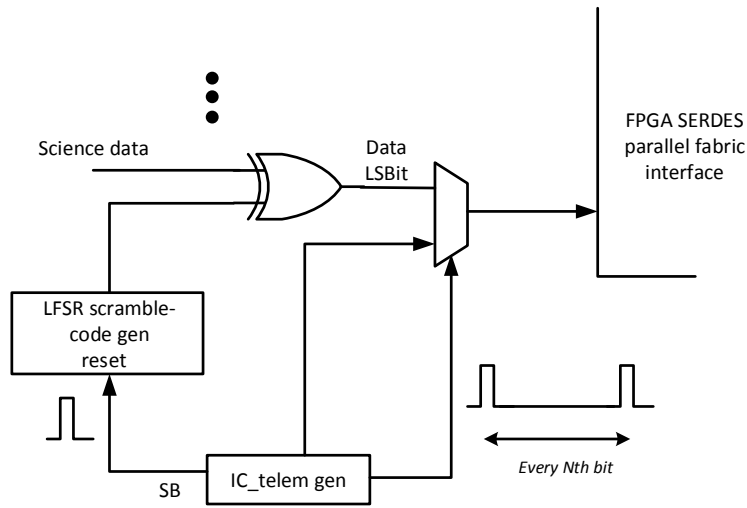

*Figure 9-2  Integration of the IC_telem protocol of Figure 4-20 in a data-replacement fashion with select LSBits of digitized science data.*

The **advantages** of Architecture A are:

- There is no separate fiber required for IC signaling using the IC_telem, or similar, protocol.
- If IC_telem provides framing and proxy error detection as described above, the entire link bandwidth can be used for digitized science data, optimizing link use.

- Since the science data delay is known via round-trip delay counting using the IC_telem SB bit mentioned previously, there is no need to send a TE to the antenna for the purposes of data time marking[36] since such can be done internally in the central FPGA after the ReSampler.
- The IC_telem frame can be used to convey the current Walsh state of the receiver to the CBF and any other antenna real-time information, with appropriate choice of "Npr" (*Figure 4-20*), N (above figure), and the clock frequency at the FPGA SERDES parallel fabric interface, to achieve the required Walsh state timing accuracy.
- The ALMA CTR (H-maser) can be located at the OSF, a more attractive site for both its location and for its maintenance.
- It is possible, TBC, that the ReSampler FIFO and interpolator could be integrated with, and indeed part of, the wideband bulk delay and 1st-stage coarse filterbank (which contains a poly-phase FIR filter), such as in the correlator/beamformer (CBF) proposed in [16].

The **disadvantages** of Architecture A are:

- The COTS PacketLight equipment listed uses a separate fiber for both transmit and receive, although likely this or other equipment providers could use a single fiber for duplex operation if found necessary. Similarly for the 400G TRx between the antennas and the AOS transponder as well as between the OSF transponder and the CBF. Separate fibers could be an IC performance limitation, which can only be evaluated with empirical testing.
- Full-duplex communications is required throughout, which is a waste of bandwidth for the reverse direction where the antenna has no need for its own data. The net cost of this may be small since COTS equipment (e.g. fiber TRx modules) is normally all full-duplex. For example, budgetary pricing for the PacketLight transponder equipment shown in Figure 9-1, for up to 128 antennas, without TRx modules to/at the antennas or the CBF is ~$1.2M.

## 9.2   Architecture B

A functional block diagram of Architecture B is shown in Figure 9-3, with a more detailed end-to-end block diagram shown in Figure 9-4. This architecture is a departure from the typical implementation of Figure 2-2 in that IC signaling is de-coupled from the digitized science data and therefore each's implementation has no bearing on the other. This has a potential performance advantage over Architecture A since COTS BiDi fiber-optic transceiver modules, with reaches in excess of 100 km are low-cost and readily available, depending on speed. A notional 625 Mbps serial data rate[37] is shown, which means that in the receiving FPGAs, the Rx CDR PLL-recovered clock can, depending on implementation, be used directly to drive the tracer DDS at a rate that can be handled by the FPGA. However, a higher or lower rate could be used with the required tracer DDS clock derived from the Rx CDR PLL-recovered clock.

---

[36] But, of course, the TE must be sent to the antenna for Walsh phase switching purposes, using some mechanism, which could be via the return path or some other mechanism.

[37] The FPGA SERDES channel may not support a data rate this low. If not, external SY87701 CDR PLL chips may be used instead (not shown in the diagram.)

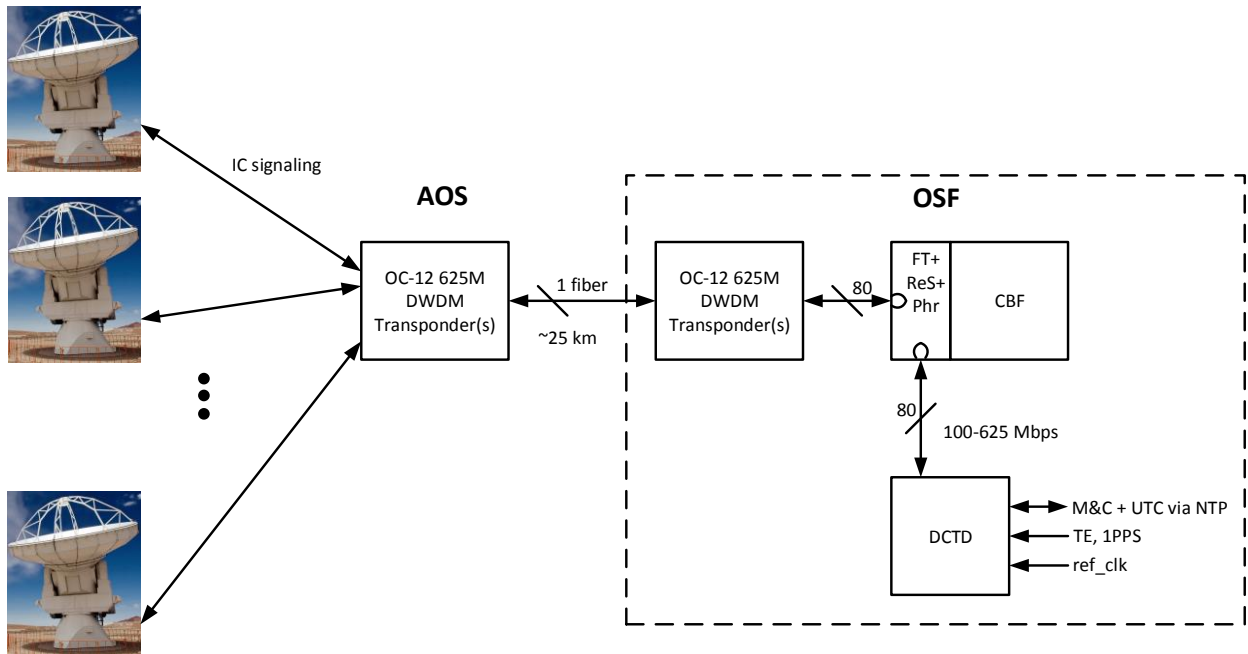

*Figure 9-3  Architecture B.  The IC timing link is de-coupled from the science data; like Architecture A, the Frequency Tracker and ReSampler are located at the OSF in front of the CBF.  Data flows to the CBF via the DTS, not shown here.*

The **advantages** of Architecture B are:

- Overall, very little return-path link bandwidth is needed, optimizing the number of fibers required for the OSF-to-AOS timing link, indeed only 1 fiber for all 80 antennas is needed, as shown in Figure 9-3, using multiple PacketLight PL-1000TE-8 and PL-300 units at each end.  This could reduce DTS equipment costs by up to a factor of 2 since no full bandwidth return DTS is needed.
- BiDi, single-fiber connections are used throughout, improving the match between the Tx and Rx directions for the round-trip phase measurement.
- The DTS need not preserve absolute timing across the link, allowing more freedom in choice of COTS equipment for its implementation, for example central-site routing via COTS packet switches.
- Similar to Architecture A, the ReSampler FIFO and interpolator could be integrated with the wideband bulk delay and 1st-stage filterbank of the CBF, TBC.

The **disadvantages** of Architecture B are:

- Since the science data delay from the antenna to the CBF is no longer known, the TE needs to be inserted into the DTS stream at the antenna.  This is a minor disadvantage over Architecture A, where time-marking of the data itself did not need to be embedded with it at the antenna.

When comparing Architecture A and B, the choice for ALMA is likely Architecture B for a) better IC timing Tx and Rx matching and b) to allow DTS design and implementation to be independent of IC timing needs.

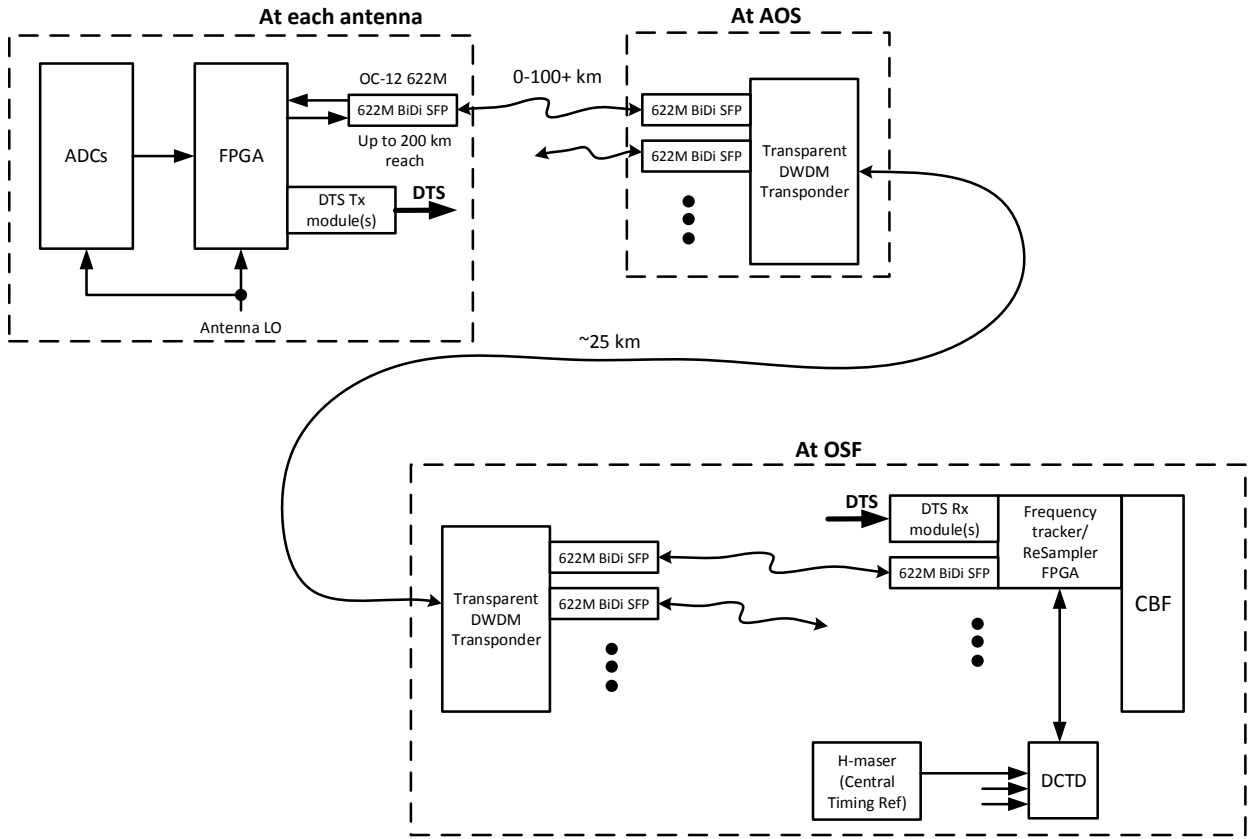

*Figure 9-4  More detailed end-to-end block diagram of Architecture B.*

## 9.3   Architecture C

Recall from section 6 that the Frequency Tracker can be provided its reference clock via a round-trip-measured phase method from a central DCTD unit, and that the fiber length from the DCTD to the Frequency Tracker can be anything, with the effect of such on a LO stability requirements governed by equations (1) and (2).   Recall also that LPFF and ReSampler processing is all discrete time and therefore can be performed at any appropriate location.   Putting these two together leads to Architecture C, shown in Figure 9-5:

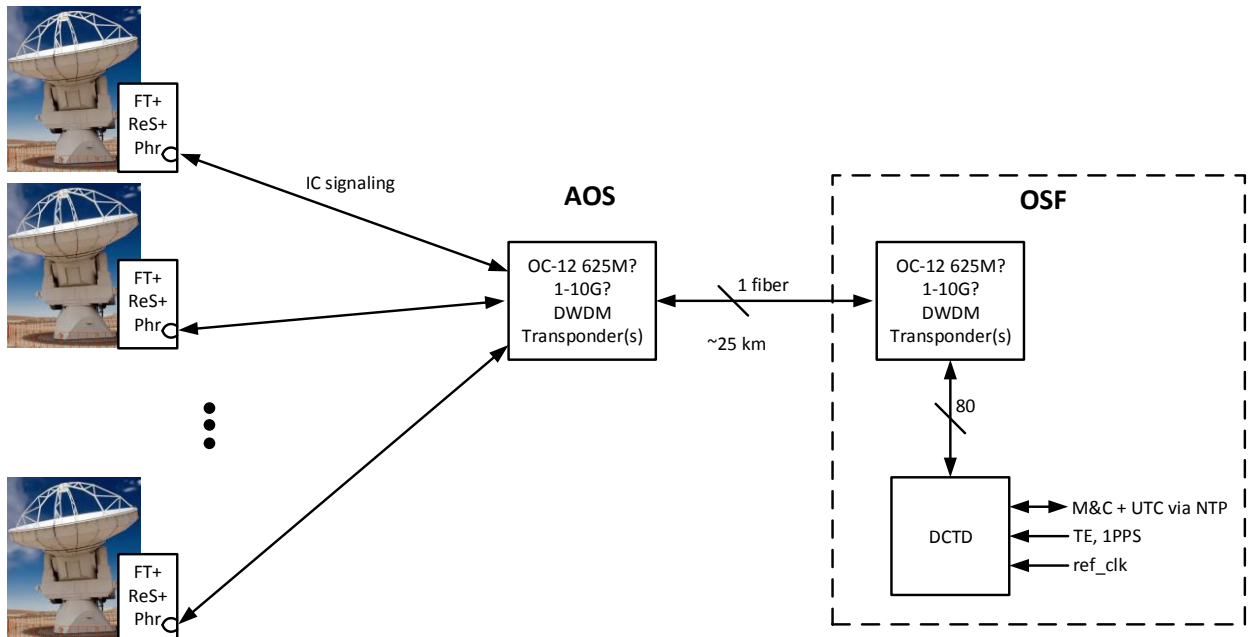

*Figure 9-5  Architecture C.  In this architecture, all Frequency Tracking and ReSampling operations are pushed out to the antenna, with the DCTD providing each instance with its reference clock, with link instabilities corrected at the antenna via a round-trip measured phase correction.*

The **advantages** of Architecture C are:

- Frequency Tracker and ReSampler processing is at the antenna and can be integrated with the FPGA(s) there, likely required anyway for interface between the ADCs and the DTS as is often the case using, for example, JESD204x as the protocol between the ADC and the FPGA and a COTS protocol such as 100GbE for the DTS.
- Since the ReSampler is at the antenna, the number of digitized sample bits feeding it from the ADC does not, particularly, have to be pruned, although ReSampler interpolator implementation resources are also a limiting factor in this regard (see section 4.4.)
- This has a more traditional topology in the sense that it starts to look like a separate clock and timing solution, with the DTS and correlator unaware of, and not caring about, its implementation.

The **disadvantages** of Architecture C are:

- The Frequency Tracker and ReSampler are located at the antenna, driving up power dissipation possibly by 200 W or more there.  SEU (Single Event Upset) events[38] are going to be more prevalent at the AOS than the OSF (due to altitude) and a larger FPGA will naturally be more susceptible to SEUs.  However, modern FPGAs have built-in single SEU upset detection and correction and multiple event detection and this significantly mitigates or may even eliminate this concern.

---

[38] For FPGAs this means that a programmed interconnect, which defines logic function, can have its setting changed, causing unpredictable erroneous logic behaviour.

- Since the ReSampler is at the antenna, re-quantization before transport via the DTS needs careful consideration. This may need to be one more bit than the ENOBs of the ADC so as not to induce an unwanted sensitivity loss, driving up the DTS and correlator ingest data rate.
- (Timing) link-loss fail-over, as previously described, may not be possible since losing the link means losing the reference clock at the antenna.

The choice between Architecture B and C is not obvious and depends on evaluation of the impact of the advantages and disadvantages, considering both ADC and DTS implementations, as well as antenna power and susceptibility to SEUs.

## 9.4 Architecture D

This architecture is a further simplification wherein the DCTD and therefore ALMA CTR, is located at the AOS, eliminating the DWDM equipment at both locations. Architecture D is shown in Figure 9-6, and now closely parallels a traditional clock & timing architecture, except it is now, essentially, all-digital.

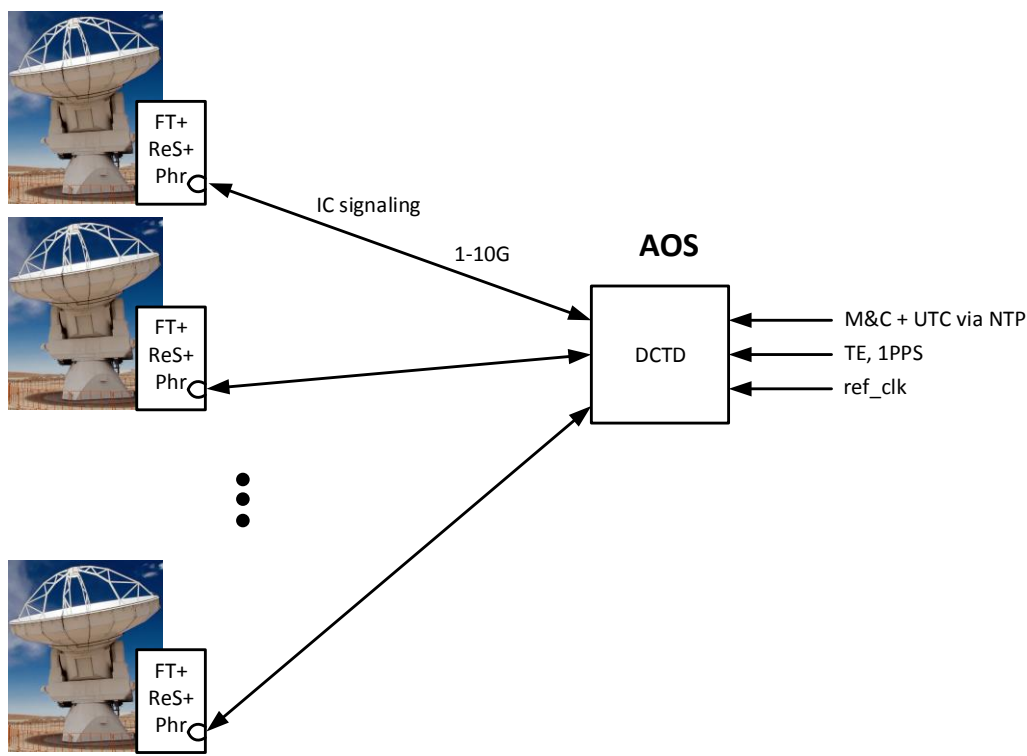

*Figure 9-6  Architecture D.  All IC equipment is located at the AOS and antennas.*

The **advantages** of Architecture D (over C) are:

- All DWDM equipment for IC signaling has been removed, simplifying the link between the DCTD and the antennas, reducing the possibility of performance limitations.
- The ALMA CTR/H-maser stays at the AOS, providing a path for a more seamless transition between the current system and ALMA 2030 systems based on the IC method. As antennas are upgraded to use IC, their output data content[39] to the CBF via the DTS, except for obviously different data rates and DTS implementation, doesn't change.

- The fall back circuit of Figure 8-3 can be used.

The **disadvantages** of Architecture D (over C) are:

- The ALMA CTR/H-maser needs to remain at the AOS, at least during the transition period between existing and ALMA 2030 systems. However, once the transition is complete, Architecture C could easily and quickly be put in place, provided of course that its performance has been verified.

## 9.5 Architecture E

This architecture is something of a merger of Architecture B and D, and possibly A if IC signaling is integrated with the DTS between the AOS and the antennas.

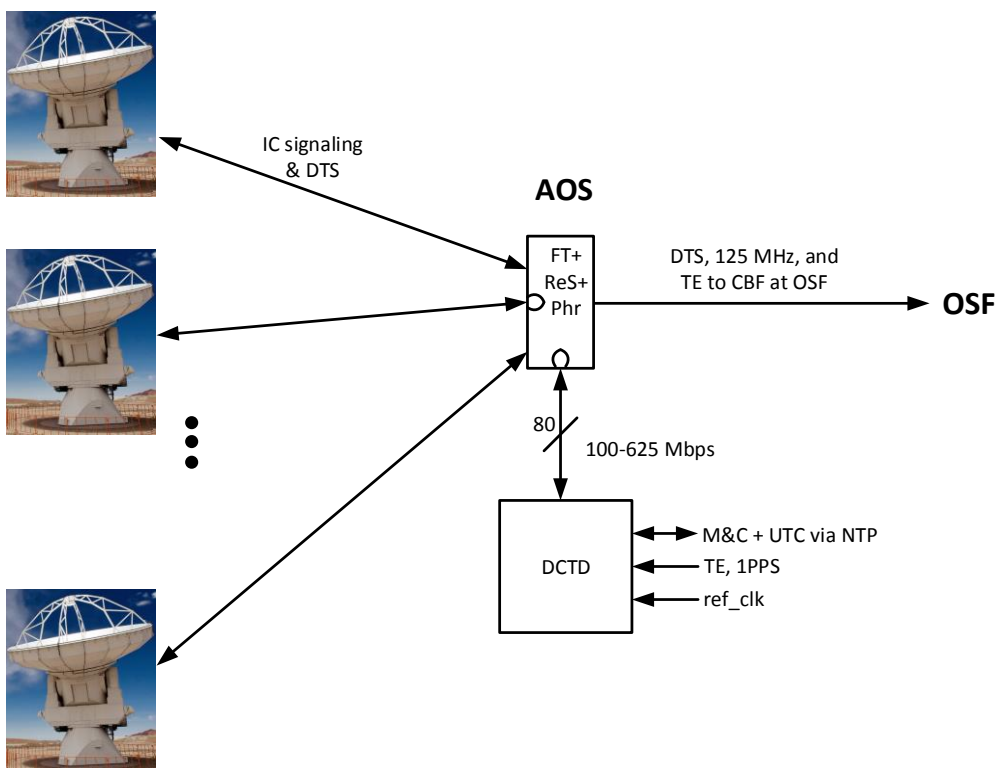

*Figure 9-7 Architecture E. All IC equipment is located at the AOS and the antennas, however with less complexity and power at the antennas than Architecture D.*

The **advantages** of Architecture E are:

- Less complexity, hardware, and power at the antennas than D.
- The fall back circuit of Figure 8-3 can be used.

The **disadvantages** of Architecture E are:

---

[39] Meaning to say that TE marking in the data stream, with a fixed and a priori established number of samples between TE marks, remains c/w Architecture A.

- The ALMA CTR/H-maser needs to remain at the AOS, although it and the DCTD could be moved to the OSF possibly with local reference clock distribution at the AOS confined such that only one (round-trip phase corrected) DCTD fiber connection to the AOS is needed (i.e. DCTD round-trip measured correction phase applies to all Frequency Trackers), although such is not necessary.

# 10 Risk analysis

As with any approach, there are risks.  This section includes a table of risks, their estimated severity, probability, impact, contingency, and probability after implementing the contingency, for all known risks, in descending order of probability.  It should be clear from this analysis, and from the evidence and theory of operation discussed in the body of this document, that the risk of an incoherent clocking approach for ALMA not working is exceedingly small if not zero.  Risks that have, in the author's opinion, zero probability are also included to indicate that they have been considered.

| Risk | Severity | Prob. | Impact or Comment | Contingency and Effort to Implement | Probability after contingency impl. |
|------|----------|-------|-------------------|-------------------------------------|-------------------------------------|
| The rtm factor in equations (1) and (2) needs to be >10. | Med | 10% | Can't achieve performance at higher/highest-band operation. | Use more stable aLO, already the case due to digital processing noise (section 4.1, allowing for rtm~=40.) Along with this, reduce $f_{c\_LPFF}$ and increase ReS FIFO depth accordingly (see section 3.) | <<1% |
| FPGA or other PLL(s) in the IC timing path have spurs at offset frequencies within $f_{c\_LPFF}$ of f_tr. | Med | 10% | Can't use FPGA or other PLLs anywhere in the IC timing path; 2 fibers for timing are required per antenna instead of 1. | Don't use any PLLs according to Figure 8-3. | 0% |
| "Stock" FPGA "IP catalog" dividers have low-level biases that lead to unacceptable DDS drift. | Med | 1% | Unacceptable delay and phase drift in visibility outputs. | Develop custom dividers with more bits, using more FPGA resources, reducing the error to an acceptable level.  This could take ~1 FTE-yr to implement. | 0% |
| Discrete-time phase sampling has spurs or causes beating effects. | High | <1% | Such has not been detected in any MathCad or FPGA RTL simulations thus far.  Inaccuracy in aLO frequency vs time measurements limit performance. | The tracer DDS Nbits and pinc are entirely flexible and can be adjusted if needed.  If there is a baseline effect, each antenna could use a different tracer pinc. | <<1% |

| Risk | Severity | Prob. | Impact or Comment | Contingency and Effort to Implement | Probability after contingency impl. |
|------|----------|-------|-------------------|-------------------------------------|-------------------------------------|
| | | | | If beating occurs and can't be eliminated, it requires actual DACs and ADCs for the tracer clock-domain crossing, requiring significant effort to implement and test, but is otherwise does not kill IC as a viable method. | |
| Cross-talk between different FPGA clock domains having to do with discrete-time phase sampling cause phase instabilities that confuse aLO measurements. | Med | <1% | Whilst this will, in principle, occur at some level it is not clear how variations at frequencies ≤ $f_{c\_LPFF}$ can be coupled. This can only be determined via empirical testing or modeling with device electrical/physical layout information. | Develop an ASIC for at least the critical discrete-time phase sampling circuit, used external to the FPGA, ensuring critical clock lines are decoupled. | 0% |
| Discrete-time phase sampling is fundamentally faulty or causes a phase drift. | | 0% | All evidence, described in section 4.3, indicates this will never occur. | N/A | N/A |
| Reference clock distribution to the "Clock domain crossing point" of Figure 4-14 introduces unacceptable differential phase wander, confusing the aLO frequency measurement. | | 0% | See section 6 for description of a round-trip phase-measured and corrected method, which will have performance similar to that shown in Figure 4-13 | N/A | N/A |

| Risk | Severity | Prob. | Impact or Comment | Contingency and Effort to Implement | Probability after contingency impl. |
|------|----------|-------|-------------------|-------------------------------------|-------------------------------------|
| Actual FPGA logic operation does not match RTL simulations | | 0% | FPGA synthesis tools have decades of development and this situation now never occurs | N/A | N/A |
| "Unknown unknowns" | | 1% | Unknown. Judged to be low since extensive FPGA RTL simulations with quasi-continuous fiber delay modeling at extreme levels beyond what is expected for actual fiber reveals no issues. | Unknown | Unknown |
| | | | | | |

# 11 Review and conclusions

A review of key advantages of the IC approach for ALMA is as follows:

- Extending ALMA baselines to 200+ km, including for Band 10.
- Virtually no restriction in the number of sub-arrays or LO tunings. Each antenna's tunings can be set independently limited only by the agility of its frequency synthesizers.
- Spurs and low-level spectral copies from interleaved digitizers don't correlate, improving the fidelity of visibility data products. The amount of de-correlation depends on aLO absolute frequency tolerance and therefore statistical frequency differences across antennas.
- All COTS digital devices/electronics such as FPGAs and lowest-cost COTS fiber-optics compared with specialized round-trip phase-corrected photonics methods.
- Fiber routed to antennas does not need to be thermally or physically stabilized; commercial long-haul fiber routing methods can be used, e.g. aerial fiber, as long as accumulated jitter is within industry jitter tolerance mask standards and the temporal delay behaviour of both directions, within the passband of the LPFF—typically ~2.5 Hz to 25 Hz—are sufficiently-well matched.
- Multiple possible architectures to choose from provide flexibility for ALMA 2030 system design including distribution or consolidation of IC processing blocks.

A review of key points of the IC approach is as follows:

- Each antenna develops and uses all of its frequencies, including for the digitizers, from its own free-running independent antenna LO (aLO.)
- Over a full-duplex digital fiber-optic connection, the aLO is measured relative to a common central LO (cLO), at a central site or even at the antenna removing, with round-trip measured phase and digital filtering, fiber delay effects on the measurement.
- Science data, digitized at a rate locked to the aLO, is re-sampled/interpolated to the common cLO clock domain before correlation and beamforming.
- Since clock-steering is not used, fiber reach is determined by aLO stability and observing frequency—the longer the reach and/or the higher the observing frequency, the more stable the aLO must be, however its absolute frequency and drifts of it, are not of concern.

A review of the state of IC research is as follows:

- Theory of operation, defined by equations and illustrated with detailed block diagrams in this document, has been verified with discrete-time mathematical modeling, FPGA RTL code simulations of key processing blocks, and end-to-end laboratory demonstrator FPGA RTL simulations including with a temporally-varying quasi-continuous-time fiber delay model.
- Discrete-time mathematical modeling of IC processing performance indicates that ALMA Band 10 1 THz operation is feasible. aLO stability requirements for this extreme case were developed; a potential COTS device is the NEL Frequency Controls 2101a OCXO (https://nelfc.com/ocxo/index.html) with ADEV~=2e-13 at Tau=0.1 sec
- Discrete-time phase sampling theory has been explained; well-known methods of memory sharing and clock-domain crossing of digital signals is employed in its implementation. An early "first LED test" FPGA hardware test was performed. This test includes discrete-time phase

**National Research Council Canada**    **Conseil national de recherches Canada**      Canada

sampling and the Frequency Tracker, yielding expected results. There is judged to be no risk that discrete-time phase sampling won't work.

- The potential risk of FPGA Rx CDR PLL and frequency synthesis PLL spurs, within the passband of the LPFF, has been identified. This risk is believed to be low but is unknown. Two fall-back positions for IC implementation have been identified, one that needs no PLLs in the IC timing path and so this risk is effectively retired.

- Many-bit "stock" digital dividers from FPGA "IP catalogs" are used in two key areas—biases here could cause systematic antenna delay and phase drift. There is no indication from the documentation or simulations that there is a bias, however, this risk can be mitigated and very likely retired using custom-developed dividers that use unbiased rounding methods. Even if some low-level bias remains, if it can be small enough, it will be inconsequential.

- The outgoing and return match of 2-fiber full-duplex connections could introduce aLO measurement errors. If this turns out to be the case, COTS "BiDi" fiber-optic transceivers, using a single fiber for both directions and with up to 120 km reach without repeating, could be used.

- A notional rtm ("round-trip multiplier") factor of 10 has been used in analysis and simulations. Any fiber round-trip phase measurements on timescales < rtm times the round-trip delay are not used in Frequency Tracker fiber delay compensation and filtering. If needed, rtm can be substantially increased if a factor of 10 is found not to be sufficient. Increasing rtm to ~40 naturally occurs anyway when Frequency Tracker digital noise is considered for Band 10 operation.

The purpose of this memo has been to present IC theory, the state of IC research, and implementation architectures for ALMA to convince the reader that IC is a viable approach for an ALMA 2030 clock and timing solution.

After these presentations and risk identification, analysis, and mitigation/retirement options, it is hoped that the reader is convinced that the chance that IC would *not* work for ALMA is small, if not zero.

**National Research Council Canada**   **Conseil national de recherches Canada**

Canada

# 13 Appendix – Derivation of the ReSampling DDS' pinc equation (5)

The general DDS equation relating its synthesized output frequency $f_{dds}$ to its phase increment $pinc_{dds}$, with $nb_{dds}$ bits, clocked at a frequency of $f_{ck\_dds}$:

$$f_{dds} = \frac{pinc_{dds}}{2^{nb_{dds}}} \cdot f_{ck\_dds} \qquad \text{(A1)}$$

Clocking the tracer DDS at a frequency divided-down by a factor of $R_{a\_t}$ from the ADC clock:

$$f_{ck\_tr\_dds} = \frac{f_{ck\_adc}}{R_{a\_t}} \qquad \text{(A1a)}$$

Substituting (A1a) into (A1) for the tracer DDS output frequency $f_{tr\_dds}$:

$$f_{tr\_dds} = \frac{pinc_{tr\_dds}}{2^{nb_{tr\_dds}}} \cdot \frac{f_{ck\_adc}}{R_{a\_t}} \qquad \text{(A1b)}$$

We want the ReSampling DDS of Figure 4-17 to synthesize the difference between the nominal ADC (digitizer) clock rate and the actual measured rate. The "nominal" rate is where the ADC's clock is frequency-locked to the common LO, cLO.

Additionally, the output of the LPFF is f_tr_diff, which is the difference (in Hz) between the actual tracer frequency that is being measured by the Frequency Tracker and the nominal tracer frequency, which is its frequency if it were frequency-locked to the cLO.

Substituting the difference between the ADC clocking frequency and its nominal, $f_{ck\_adc} - f_{ck\_adc\_nom}$ for $f_{ck\_adc}$, and $f_{tr\_diff}$ for $f_{tr}$ into equation (A1b), and we get:

$$f_{tr\_diff} = \frac{pinc_{tr\_dds}}{2^{nb_{tr\_dds}}} \cdot \frac{(f_{ck\_adc} - f_{ck\_adc\_nom})}{R_{a\_t}} \qquad \text{(A1c)}$$

Which simplifies to (A1b) since $f_{tr\_diff} = f_{tr} - f_{tr\_nom}$ and $f_{tr\_nom} = (pinc_{tr\_dds}/2^{nb\ tr\_dds}) \cdot f_{ck\_adc\_nom}/R_{a\_t}$ from (A1).

As noted above, $f_{ck\_adc} - f_{ck\_adc\_nom}$ is the frequency we want the ReSampling DDS to synthesize, $f_{ReS\_dds}$. Substituting $f_{ReS\_dds} = (f_{ck\_adc} - f_{ck\_adc\_nom})$ into (A1c) and re-arranging:

National Research Council Canada    Conseil national de recherches Canada

Canada

$$f_{ReS\_dds} = \frac{R_{a\_t} \cdot 2^{nb_{tr\_dds}}}{pinc_{tr\_dds}} \cdot f_{tr\_diff} \qquad \text{(A2)}$$

Then, from (A1), for the ReSampling DDS:

$$f_{ReS\_dds} = \frac{pinc_{ReS\_dds}}{2^{nb_{ReS\_dds}}} \cdot f_{ck\_ReS\_dds} \qquad \text{(A3)}$$

Finally, substituting (A2) into (A3), solving for the ReSampling DDS' pinc, $pinc_{ReS\_dds}$ and simplifying, we get:

$$pinc_{ReS\_dds} = \frac{2^{(nb_{ReS\_dds}+nb_{tr\_dds})} \cdot R_{a\_t} \cdot f_{tr\_diff}}{f_{ck\_ReS\_dds} \, pinc_{tr\_dds}} \qquad \text{(A4)}$$

Which is exactly equation (5) on p. 29.

National Research    Conseil national
Council Canada    de recherches Canada

Canada